\documentclass[]{aastex7}

\hypersetup{urlcolor=blue}

\begin{document}

\title{TNO colours provide new evidence for a past close flyby of another star to the Solar System}

\newcommand{\noop}[1]{}
\newcommand{\okina}{\textquoteleft}
\newcommand{\OU}{{\okina}OUMUAMUA}
\newcommand{\Ou}{1I/{\okina}Oumuamua}
\newcommand{\borisov}{2I/Borisov}
\newcommand{\ME}{M$_\Earth$\,}
\newcommand{\RE}{R$_\Earth$\,}
\newcommand{\paperone}{\mbox{paper 1}}
\newcommand{\papertwo}{\mbox{paper 2}}
\newcommand{\be}{\begin{eqnarray}}
\newcommand{\ee}{\end{eqnarray}}

\newcommand{\Myr} {\mbox{$~\text{Myr}$}}
\newcommand{\AU} {\mbox{$~\text{AU}$}}
\newcommand{\pc} {\mbox{$~\text{pc}$}}
\newcommand{\pcsqu} {\mbox{$~\text{pc}^{-2}$}}
\newcommand{\pccu} {\mbox{$~\text{pc}^{-3}$}}
\newcommand{\MSun} {\mbox{$M_{\odot}$}}
\newcommand{\MJup} {\mbox{$~M_{Jup}$}}
\newcommand{\Mcluster} {\mbox{$~M_{\text{cl}}$}}
\newcommand{\Mstars} {\mbox{$~M_{\text{stars}}$}}
\newcommand{\Mgas} {\mbox{$~M_{\text{gas}}$}}
\newcommand{\rhoM} {\mbox{$~M_{\odot}~\text{pc}^{-3}$}}
\newcommand{\rhoStars} {\mbox{$~\text{stars}~\text{pc}^{-3}$}}

\graphicspath{{./}{Figures_final/}}

\author[0000-0002-5003-4714]{Susanne Pfalzner}
\affiliation{J\"ulich Supercomputing Center, Forschungszentrum J\"ulich, 52428 J\"ulich, Germany}
\email{s.pfalzner@fz-juelich.de}

\author[0009-0001-5904-1742]{Frank W. Wagner}
\affiliation{J\"ulich Supercomputing Center, Forschungszentrum J\"ulich, 52428 J\"ulich, Germany}
\email{f.wagner@fz-juelich.de}

\author[0000-0002-5540-9626]{Paul Gibbon}
\affiliation{J\"ulich Supercomputing Center, Forschungszentrum J\"ulich, 52428 J\"ulich, Germany}
\email{p.gibbon@fz-juelich.de}

\begin{abstract}

Thousands of small bodies, known as trans-Neptunian objects (TNOs), orbit the Sun beyond Neptune. TNOs are remnants of the planets' formation from a disc of gas and dust, so it is puzzling that they move mostly on eccentric orbits inclined to the planetary plane and show a complex red-to-grey colour distribution. A close stellar flyby can account for the TNOs' dynamics but it is unclear if this can also explain the correlation between their colours and orbital characteristics. Assuming an initial red-to-grey colour gradient in the disc, our numerical study finds that the spiral arms induced by the stellar flyby simultaneously lead to the observed TNOs' colour patterns and orbital dynamics. The combined explanation of these TNO properties strengthens the evidence for a close flyby of another star to the young Solar System. Our study predicts that (1) small TNOs beyond 60~au will mostly be grey, and (2) retrograde TNOs will lack the colour most common to high-inclination TNOs. The anticipated TNO discoveries by the Vera Rubin telescope will be able to test these predictions. A confirmed flyby would allow us to reveal the chemical composition of the Solar System's primordial disc.

\end{abstract}

%


\section{Introduction}
\label{sec:intro}

More than 3,000 subplanet-sized objects have been discovered beyond Neptune. These trans-Neptunian objects (TNOs) formed alongside the Solar System's planets and provide valuable insight into the Solar System's early history \citep{Gladman:2021}. Most TNOs are too faint for spectroscopic observations \citep{Pinilla:2020, Glein:2023, Emery:2024}, so their surface compositions are mainly studied through broadband photometry that reveals colours ranging from very red to blue-grey \citep{Tegler:1998, Brown:2012, Barucci:2020}.

The temperature decreases with increasing distance from the Sun, so that one would expect the TNOs' colours to correlate with this distance too. However, no such relationship has been found \citep{Jewitt:2001}. Instead, the TNOs' colours appear to be related to their dynamic characteristics. Low-inclination and low-eccentricity TNOs in the cold Kuiper belt are predominantly very red \citep{Tegler:2000,Hainaut:2002,Doressoundiram:2005,Peixinho:2008}, while TNOs in the hot Kuiper belt moving on inclined, eccentric orbits exhibit a mix of red and grey tones. Recent surveys (Outer Solar System Survey (OSSOS) \citep{Schwamb:2019}, Dark Energy Survey (DES) \citep{Bernardinelli:2023}, James Webb Space Telescope (JWST) observations \citep{Pinilla:2024}) indicate that very red TNOs are rare at inclinations, $i >$~21~° \citep{Marsset:2019} and eccentricities, $e >$~0.42 \citep{Ali:2021}. The underlying reason for these correlations is unclear.

Here, we investigate a close flyby of another star as a possible reason for the correlations between colour and dynamics. Specifically, we consider a 0.8~\MSun\ star passing at a periastron distance \mbox{$r_P =$~110~au} with an inclination \mbox{$i_P =$~70~°}. This flyby reproduces the dynamics of currently known TNOs remarkably well \citep{Pfalzner:2024a}. The question is whether this flyby can also account for the observed colour distribution of the known TNOs.

\begin{figure*}
  \centering
  
  \begin{minipage}[b]{0.47\textwidth}
    \centering
    \includegraphics[width=\textwidth]{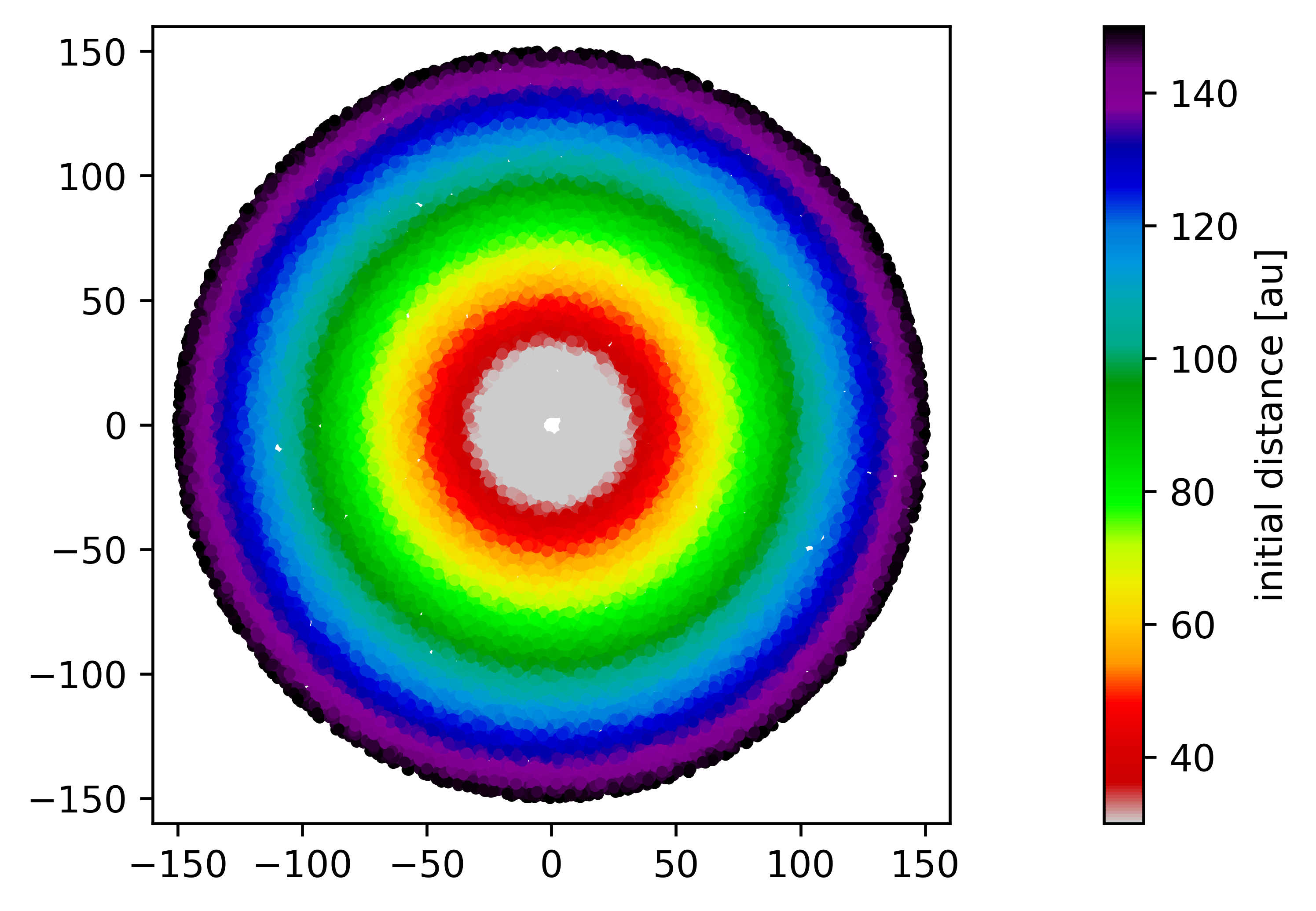}
    \textbf{(a)}
  \end{minipage}
  \begin{minipage}[b]{0.47\textwidth}
    \centering
    \includegraphics[width=\textwidth]{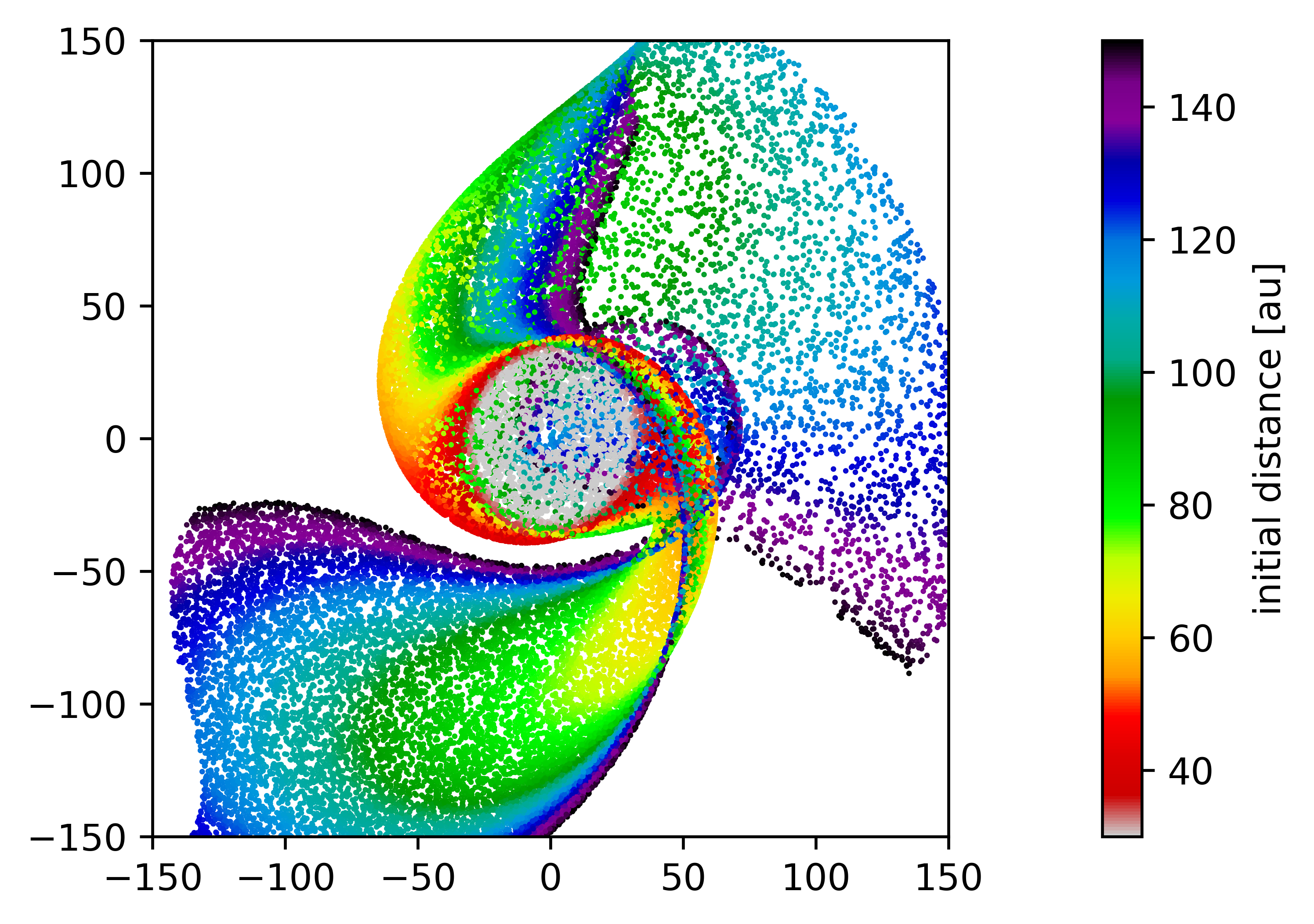}
    \textbf{(b)}
  \end{minipage}
  
  \caption{{\bf Effect of flyby.} (a) The pre-flyby colour gradient in the simulated disc is depicted by a false colour scheme representing very red to blue-grey TNOs. The same colour scheme is used throughout the paper. (b) Simulation snapshot of the flyby 128 years after periastron passage. The perturber star approached from the bottom right and already left the shown area. Disc matter is transported inwards and outwards along the spiral arms, with a fraction of the test particles injected into the planet region. Only the particles remaining bound to the Sun are shown.}
  \label{fig:flyby}
\end{figure*}

\section{Method}
\label{sec:meth}

We simulate the effect of the flyby of a 0.8~\MSun\ star on a disc modelled with 10,000 and 50,000 massless test particles initially orbiting the Sun on Keplerian trajectories. The size of the primordial solar disc is unknown. Observed sizes of protoplanetary and debris discs typically range from 100~au to 500~au \citep{Andrews_2020,Hendler_2020}. We model the effect of a flyby up to a radius of 150~au. The results for smaller-sized discs are contained as a subset of the results data. We model the disc by an initially constant test particle surface density. The reason is that the outer disc areas are better resolved with such an approach. In the second step, one can assign different masses to the particles to model the actual mass density \citep{Hall:1997,Steinhausen:2012, Pfalzner:2024a}. The Appendix contains examples for $1/r$- and $1/r^{3/2}$-dependent initial surface densities.

We assume a colour gradient in the pre-flyby disc and represent it by a rainbow colour spectrum between 30~au and 150~au as illustrated in Fig.~\ref{fig:flyby}~(a). The cold Kuiper belt objects move on near-circular orbits close to the plane. It is generally assumed that these cold Kuiper belt objects remained close to their place of origin
\citep{Parker:2010, Petit:2011,Fraser_2021}. These bodies are nearly exclusively very red in colour \citep{Thirouin:2019,Fraser_2021,Bernardinelli:2025}. Therefore, we assume that the primordial disc was dominated by very red bodies in the cold Kuiper belt region and that beyond this, objects of various shades of grey existed. Our rainbow colour representation of this situation has two functions. First, it is a temporary placeholder for the observed slope $S$ and will eventually link to the TNOs' compositions. Second, it links to the primordial disc size. One can locate which particles would be absent if the primordial disc were smaller than 150~au.

The perturber star (\mbox{$M_P =$~0.8~\MSun}) approached to a periastron distance \mbox{$r_P =$~110~au} on an inclined orbit \mbox{$i_P =$~70~°} with an argument of periastron \mbox{$\omega_P =$~80~°}. The flyby probably took place during the early phases of the Solar System in the Sun's birth cluster. In such clusters, the stellar density is about 1,000 to a million times higher than the local stellar density, and therefore, close flybys are much more common. Due to the stars in a young cluster being almost coeval, most flybys happen on nearly parabolic orbits even in dense stellar groups, such as the Orion Nebula cluster \citep{Olczak:2010}. At least 140 million solar-type stars in the Milky Way may have experienced a similar flyby in their youth \citep{Pfalzner:2024a}.

A flyby with the above parameters effectively reproduces all the known dynamic groups among the TNOs \citep{Pfalzner:2024a} apart from the resonant populations, which appear later on through interactions with Neptune. The disc's response is modelled using the REBOUND code \citep{Rein:2014} with the IAS15 integrator. This $N$-body approach is sufficient as the gas mass in the debris disc is negligible \citep[e.g., ][]{Kobayashi:2001, Musielek:2014, Pfalzner:2018}. For more detailed information on the simulation method, see \citet{Pfalzner:2024b}. The perturber's influence on test particle orbits becomes negligible approximately 12,000 years after periastron passage.

To study the long-term evolution after the flyby, we use the GENGA code \citep{Grimm:2014, Grimm:2022} with its hybrid symplectic integrator \citep{Chambers:1999}. Here, we also consider the interactions with the four giant planets of the Solar System. We limited our simulations to the first 1~Gyr after the flyby, because the computational cost remains relatively high despite the high numerical efficiency of GENGA. We find a time step of \mbox{$\Delta t =$~23~d} is sufficient to obtain the required temporal resolution.

\begin{figure*}
\centering

  \begin{minipage}[b]{0.40\textwidth}
    \centering
    \includegraphics[width=\textwidth]{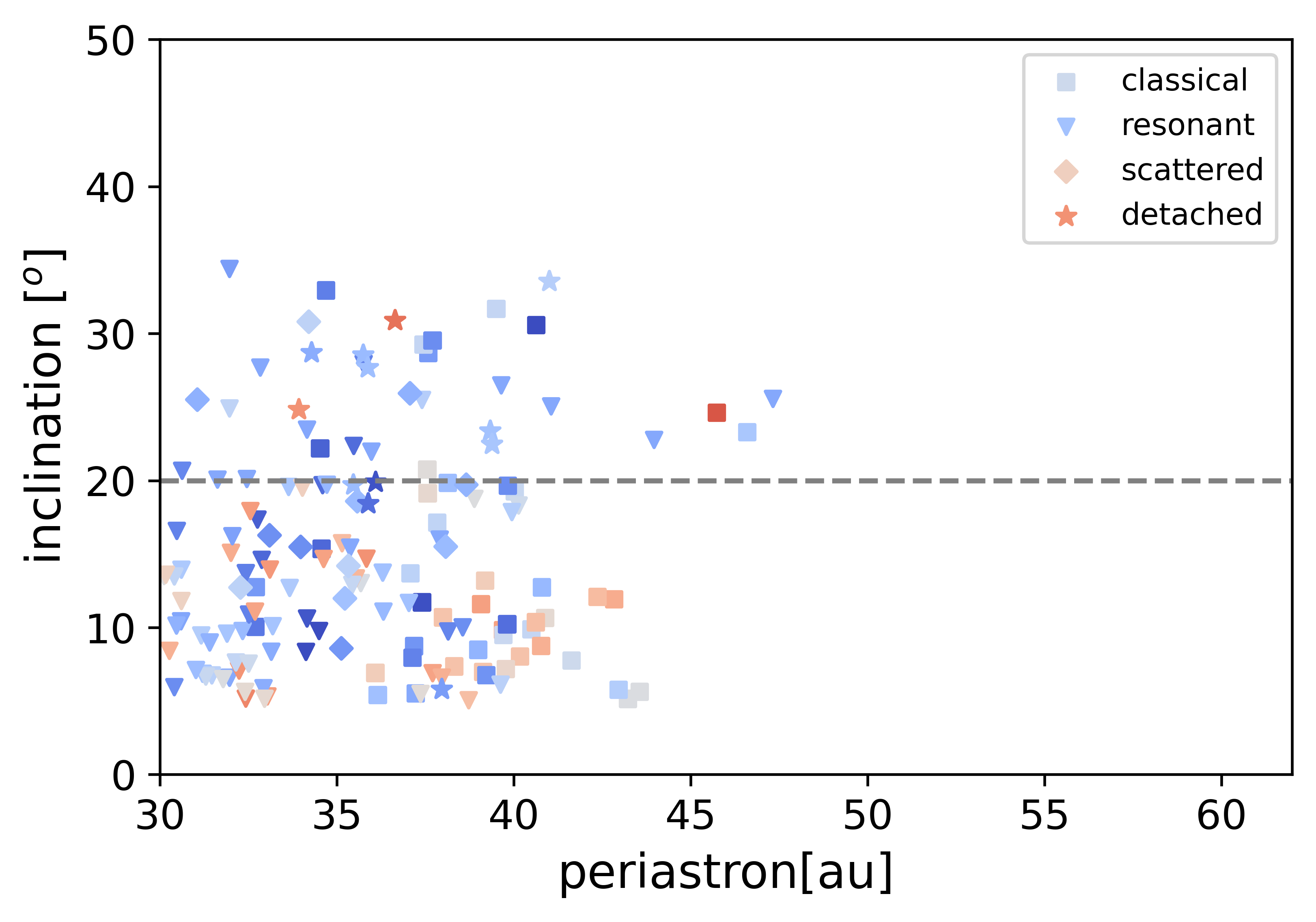}
    \textbf{(a)}
  \end{minipage}
  \begin{minipage}[b]{0.40\textwidth}
    \centering
    \includegraphics[width=\textwidth]{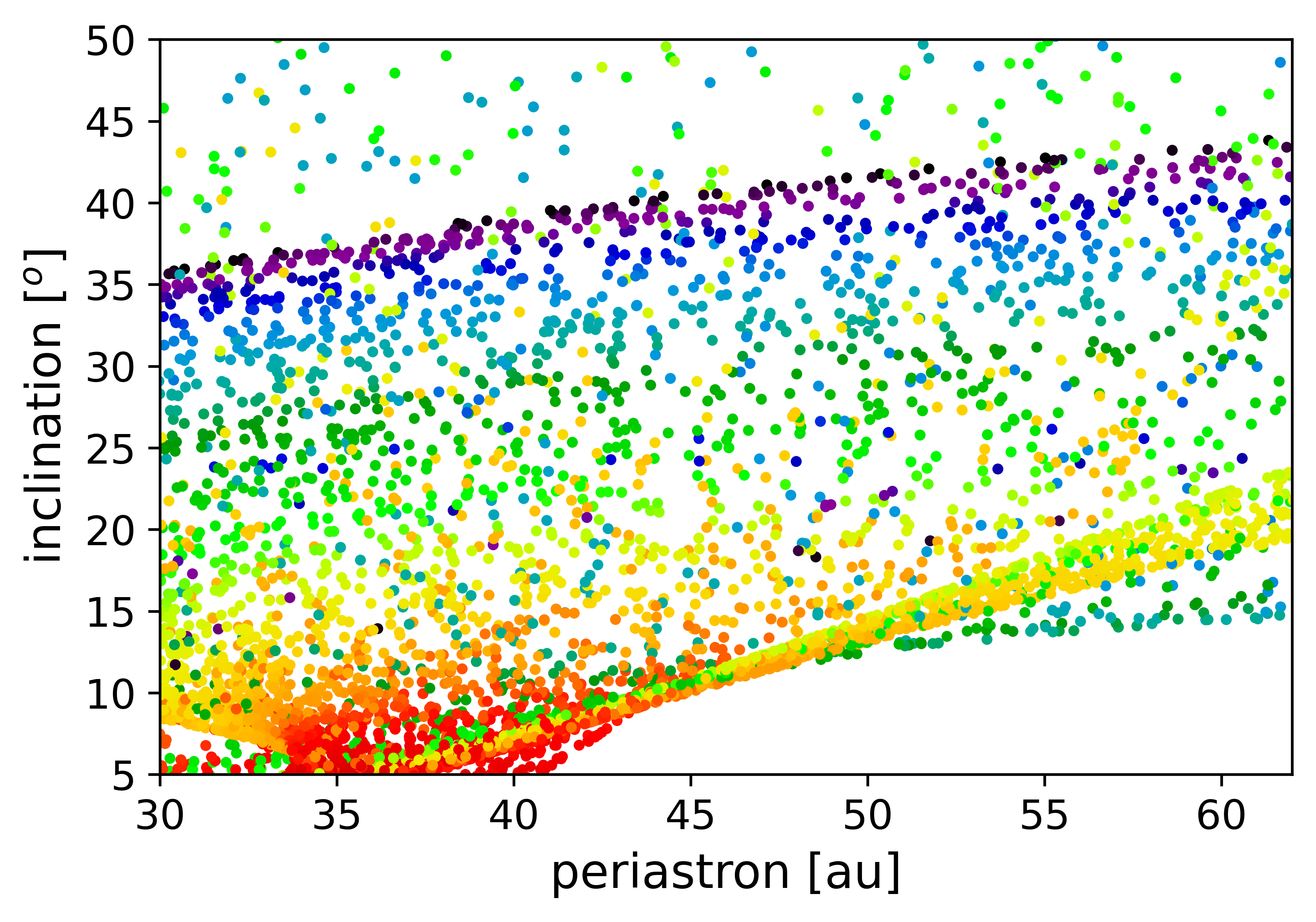}
    \textbf{(b)}
  \end{minipage}

  \begin{minipage}[b]{0.37\textwidth}
    \centering
    \includegraphics[trim=0.7cm 0cm 0.55cm 0cm, clip, width=\textwidth]{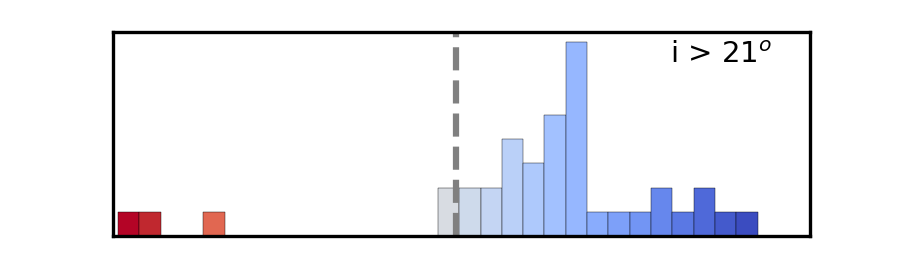}
    \includegraphics[width=\textwidth]{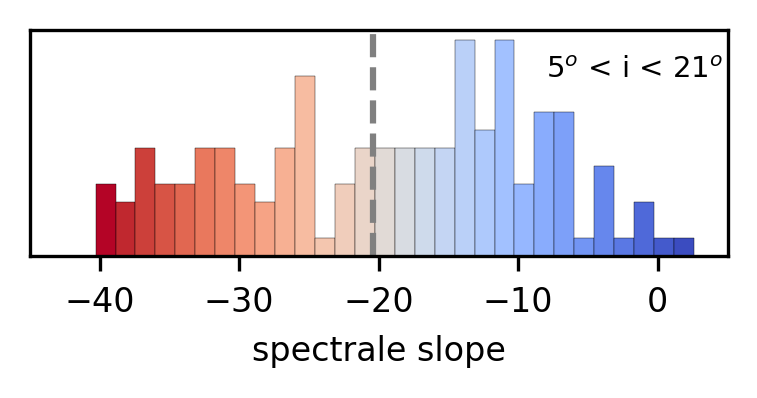}
    \textbf{(c)}
  \end{minipage}
  \begin{minipage}[b]{0.37\textwidth}
    \centering
    \includegraphics[width=\textwidth]{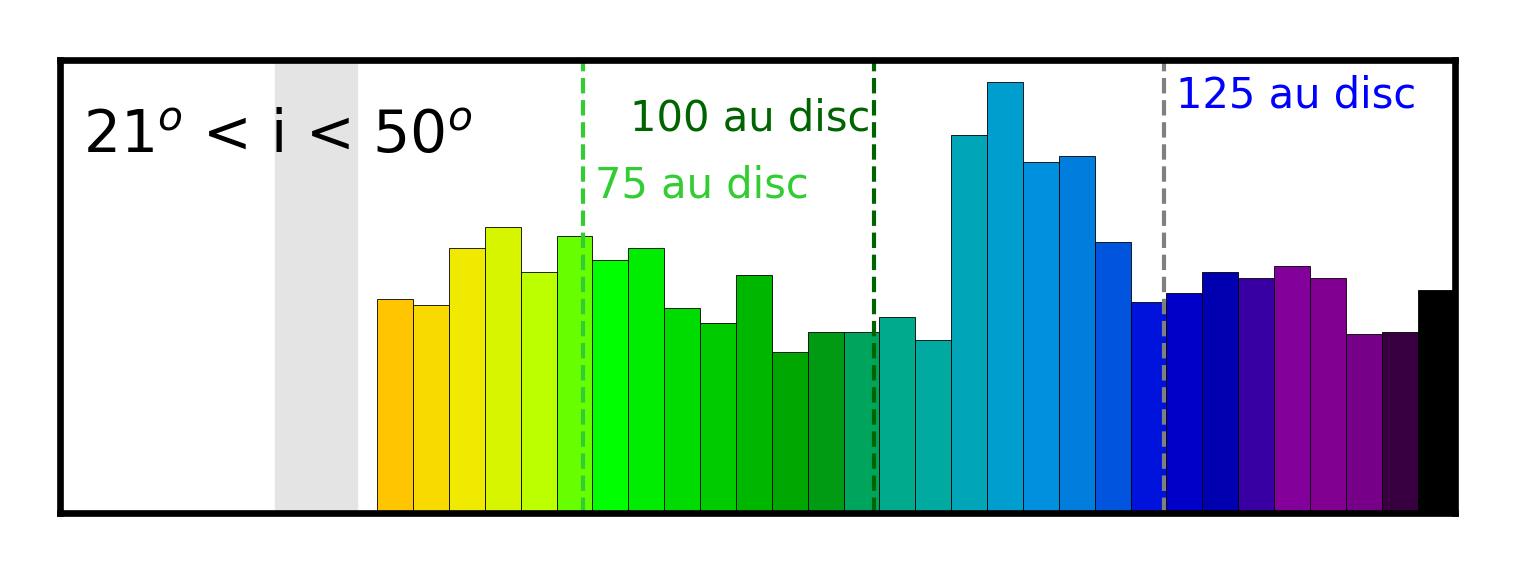}
    \includegraphics[width=1.035\textwidth]{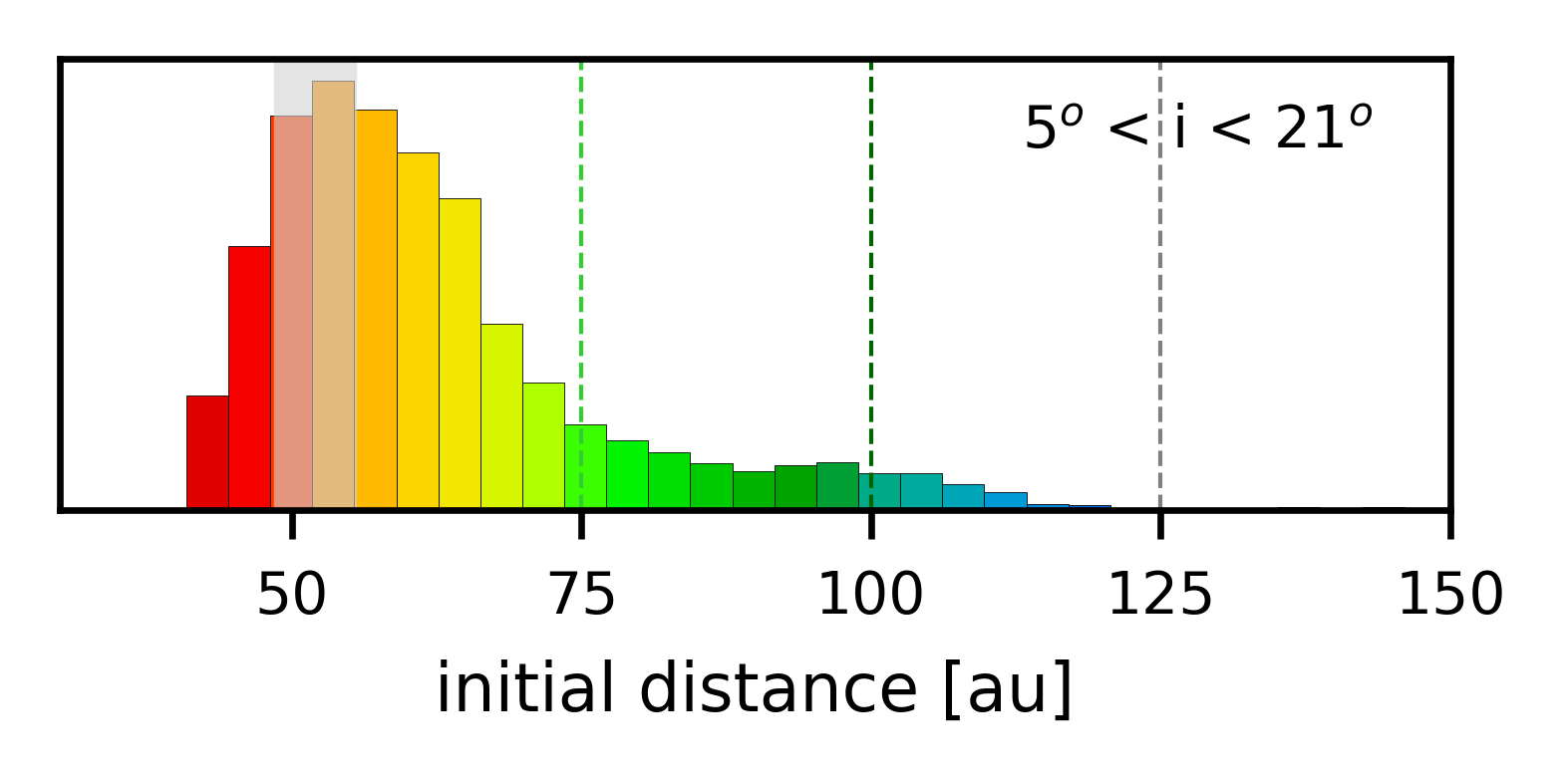}
    \textbf{(d)}
  \end{minipage}

  \begin{minipage}[b]{0.37\textwidth}
    \centering
    \includegraphics[width=\textwidth]{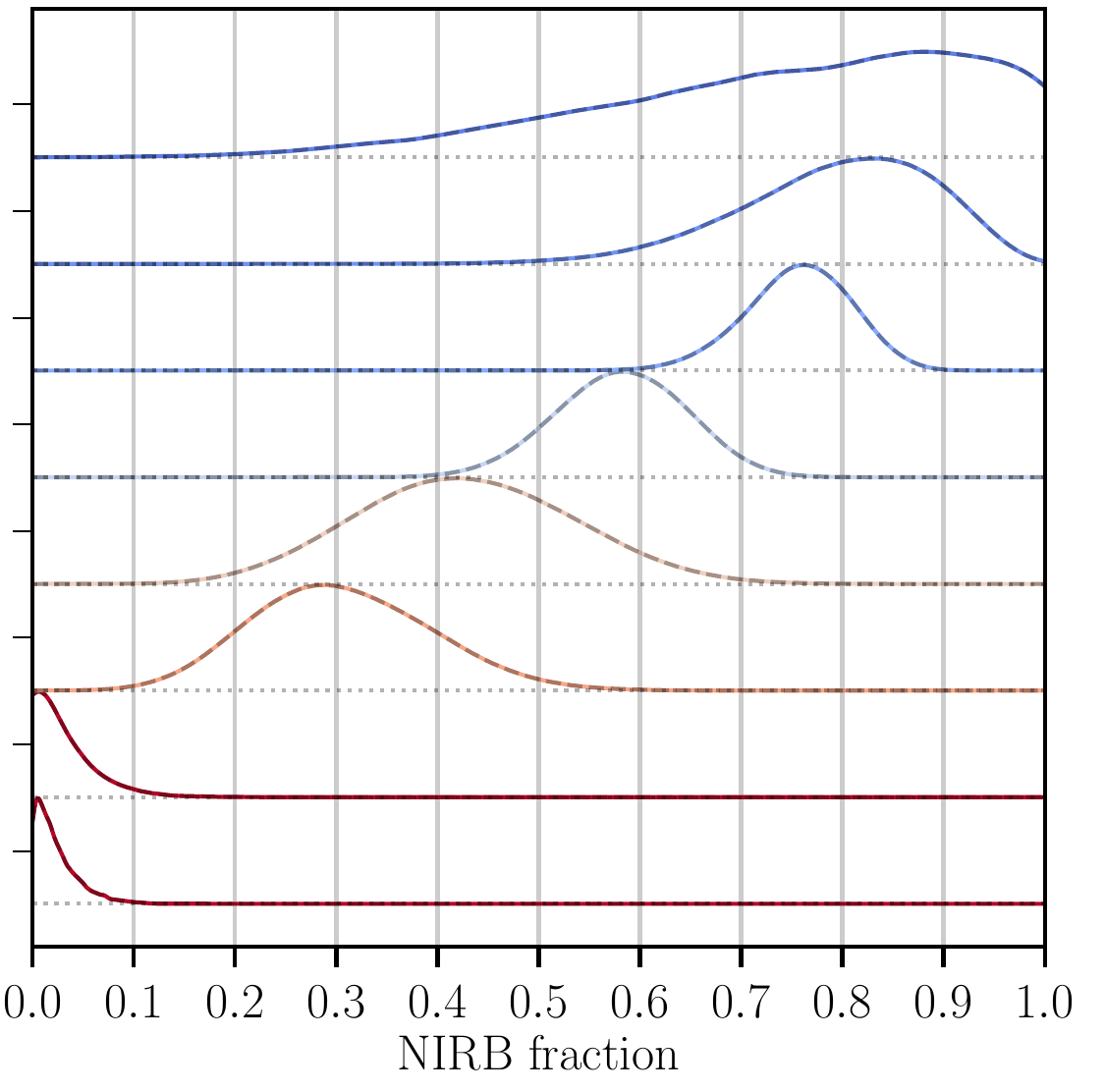}
    \textbf{(e)}
  \end{minipage}
  \begin{minipage}[b]{0.34\textwidth}
    \centering
    \includegraphics[width=\textwidth]{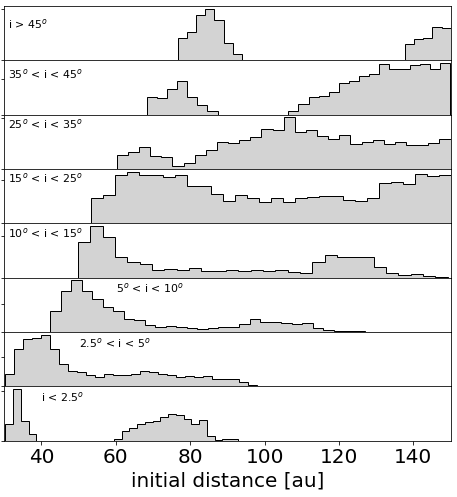}
    \textbf{(f)}
  \end{minipage}

\caption{{\bf Connection between TNO colour and inclination.} The left column illustrates the observational data of \citet{Marsset:2019} and \citet{Bernardinelli:2025}, and the right column shows the numerical results from our flyby simulation. The top row displays scatter plots of the TNOs' inclination as a function of periastron distance. The middle row shows colour distributions for the high-inclination ($i >$~21~°) and the low-inclination (5~°~$< i <$~21~°) objects, respectively. The colours in panels (b) and (d) indicate the distance to the Sun in the pre-flyby disc as defined in Fig.~\ref{fig:flyby}~(a). The colours can also be used to convert the results to smaller initial disc sizes. The vertical lines in panel (c) correspond to the grey area in panel (d) and show the border between very red and grey TNOs in the observations and simulations. For the data of \citet{Marsset:2019} we excluded the centaurs. (e) Observations of the fraction of infrared bright objects as a function of inclination bin for the classical TNOs in the DES survey \citep{Bernardinelli:2025}. (f) Simulation results of the origin distribution for the same inclination bins. Only objects with 30~au~$< p <$~60~au are considered.}
\label{fig:compare_inc}
\end{figure*}

\section{Results}
\label{sec:res}

We simulate the effect of the stellar flyby on a debris disc. We assume a colour gradient as illustrated in Fig.~\ref{fig:flyby}~(a). The colour gradient is used as a temporary placeholder for the observed slope $S$. During the flyby, some TNOs become unbound, while others are captured by the perturber. However, most remain bound to the Sun \citep{Pfalzner:2024a}. The perturber significantly alters their orbits, creating visible spiral arms due to the induced sub- and super-Keplerian velocities (Fig.~\ref{fig:flyby}~(b)). The sub-Keplerian stream moves planetesimals inwards (dark blue), whereas the super-Keplerian stream pushes particles outwards (red/yellow) \citep{Pfalzner:2003}.

Colours are primarily known for relatively close TNOs \citep{Gladman:2021}, mostly from the Kuiper belt region. However, these TNOs represent only a portion of the entire population of discovered TNOs. In addition, the real number of TNOs likely very much exceeds that of the known TNOs. For example, the number of undiscovered detached TNOs ($a >$ 47.2 au, $e >$ 0.42) is expected to exceed that of the currently known Kuiper belt population \citep{Gladman:2021}. For our comparison with the observations, we focus on TNOs with $p <$~60~au and $i <$~50~°. By contrast, we look at the extended parameter space for predictions of future discoveries. An illustration of the entire parameter space is given by Figs.~\ref{fig:flyby_entire}~(a) and (b) of the Appendix.

\subsection{Inclination correlation}
\label{sec:inc}

Figure~\ref{fig:compare_inc} illustrates the relationship between the TNOs' colours, inclinations, and periastron distances, comparing the OSSOS observations in the left column to our simulations on the right. \citet{Marsset:2019} analysed the colours of 229 TNOs on dynamically excited orbits from the OSSOS survey. They find that very red objects are rare at inclinations $i >$~21~° (see, Fig.~\ref{fig:compare_inc}~(a)). Figure~\ref{fig:compare_inc}~(c) shows the corresponding distributions. \citet{Marsset:2019} classified TNOs as red ($S >$~20.4) or grey ($S <$~20.4) based on spectral slope $S$, as indicated by the grey dashed lines. Note the inverted $x$-axes, chosen for a better comparison with the simulations.

The simulations reflect similar findings: red test particles are mainly found at low inclinations and periastron distances, suggesting that they retain more of their original dynamics. In contrast, green to blue objects dominate at higher inclinations, with red and orange particles being rare (see, Figs.~\ref{fig:compare_inc}~(b) and (d)). The red test particles correspond to the very red TNOs, whereas the other colours represent the shades of grey observed for TNOs. Red TNOs appear to be more common at periastron distances 35~au~$< p <$~45~au, although this could be affected by the low number of TNOs with $p >$~45~au. The grey bars correspond roughly to the grey dashed lines in Fig.~\ref{fig:compare_inc}~(c). The simulation data are available at \url{https://destiny.fz-juelich.de}.

Figures~\ref{fig:compare_inc}~(b) and (d) can also be used to infer the results for smaller-sized discs. In Fig.~\ref{fig:compare_inc}~(d), we indicated where the distribution would end for discs of size 75~au, 100~au, and 125~au, respectively. For Fig.~\ref{fig:compare_inc}~(b), one has to subtract the corresponding colours from the plot. We illustrate this process for a 100~au-sized system by Fig.~\ref{fig:flyby_entire}~(c) of the Appendix. In the Appendix, we also show that the result of the black of red objects at high inclinations is valid not only for the test particle distribution shown here, but also for $1/r$ (blue)  and $1/r^{3/2}$ (red) surface distributions in the pre-flyby disc (see, Fig.~\ref{fig:surface}).

The consistency of these observed trends is further supported by \citet{Bernardinelli:2025}'s analysis of 697 TNOs from DES, which confirmed the OSSOS findings. They classified TNOs into two groups: ``near-IR bright" (NIRB) and ``near-IR faint" (NIRF), revealing a systematic shift in their ratio with higher inclinations (see, Fig.~\ref{fig:compare_inc}~(e)). Our simulations (Fig.~\ref{fig:compare_inc}~(f)) also show this change, with a surprising degree of match despite the approximate colour definition. Even the slightly backwards trend towards redder colours in the 40~° to 60~° range is evident. However, here, the statistics are less robust than in the other inclination intervals.

\begin{figure*}
\centering

  \begin{minipage}[b]{0.40\textwidth}
    \centering
    \includegraphics[width=\textwidth]{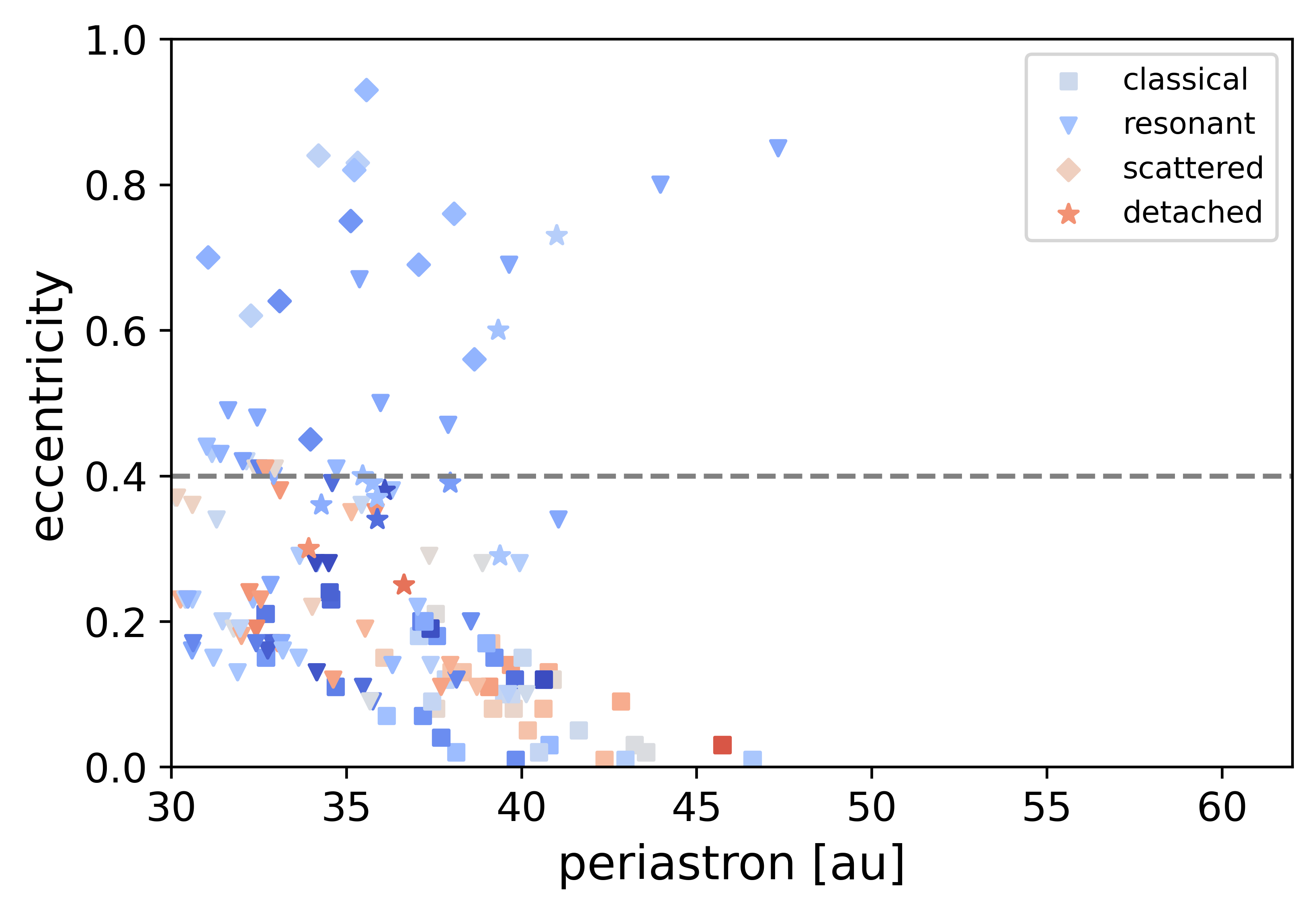}
    \textbf{(a)}
  \end{minipage}
  \begin{minipage}[b]{0.40\textwidth}
    \centering
    \includegraphics[width=\textwidth]{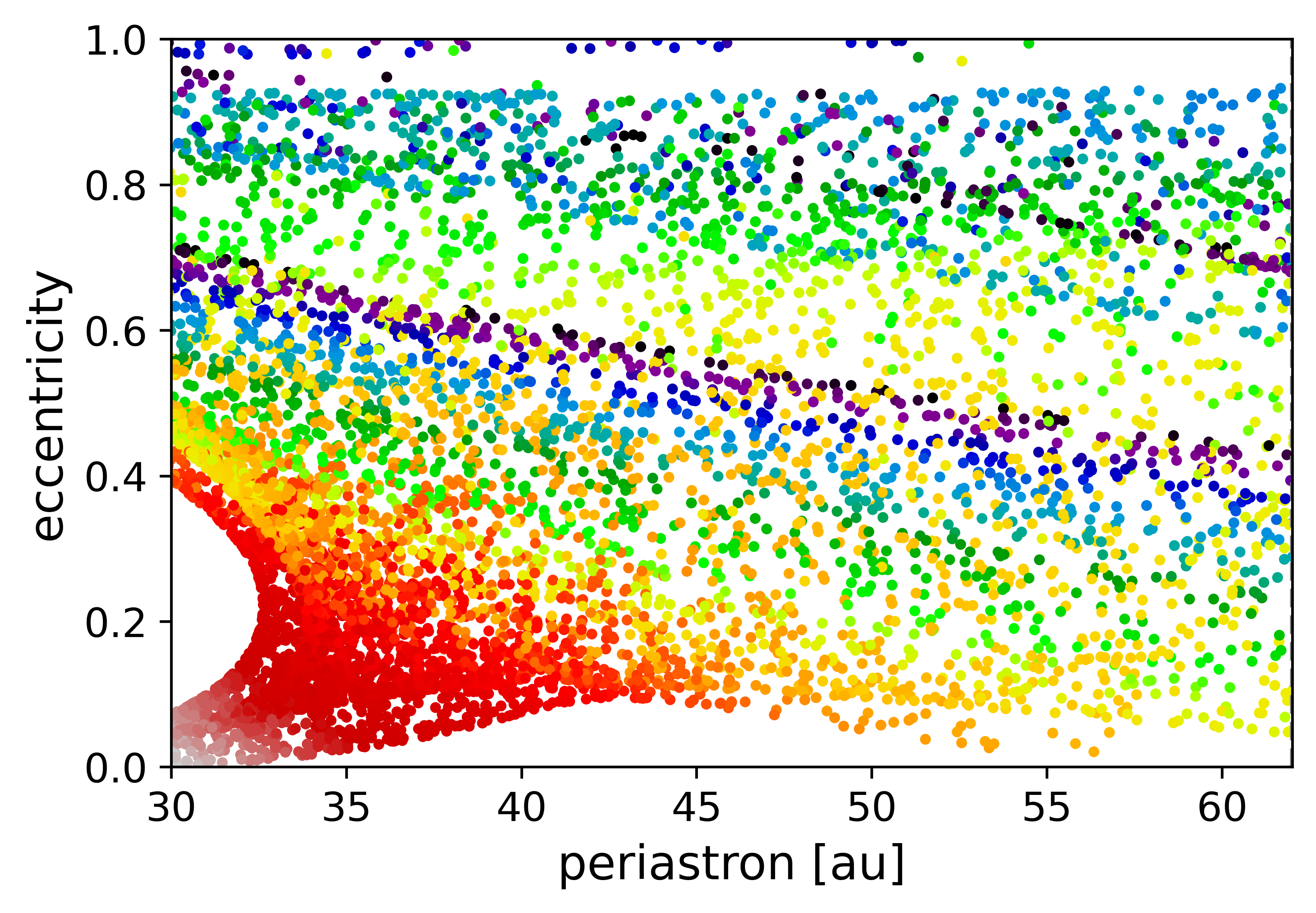}
    \textbf{(b)}
  \end{minipage}

  \begin{minipage}[b]{0.40\textwidth}
    \centering
    \includegraphics[width=\textwidth]{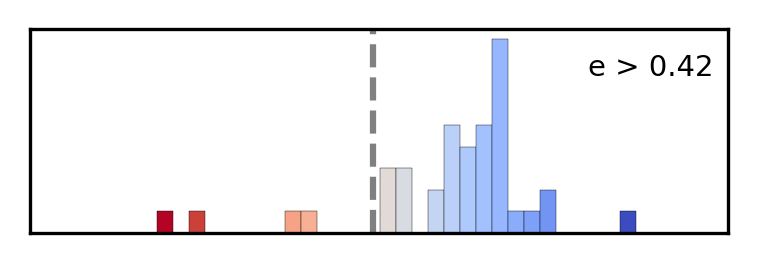}
    \includegraphics[width=\textwidth]{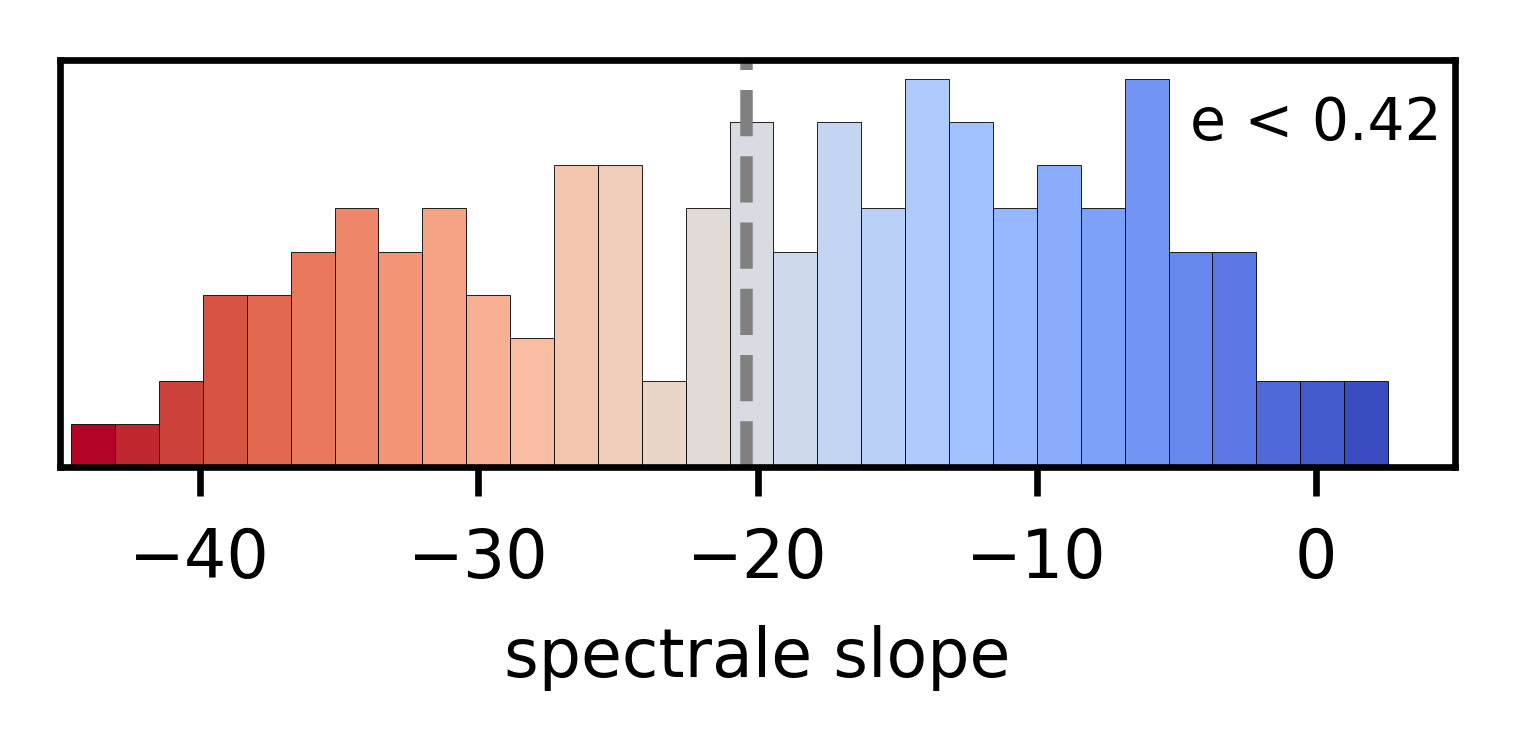}
    \textbf{(c)}
  \end{minipage}
  \begin{minipage}[b]{0.375\textwidth}
    \centering
    \includegraphics[width=\textwidth]{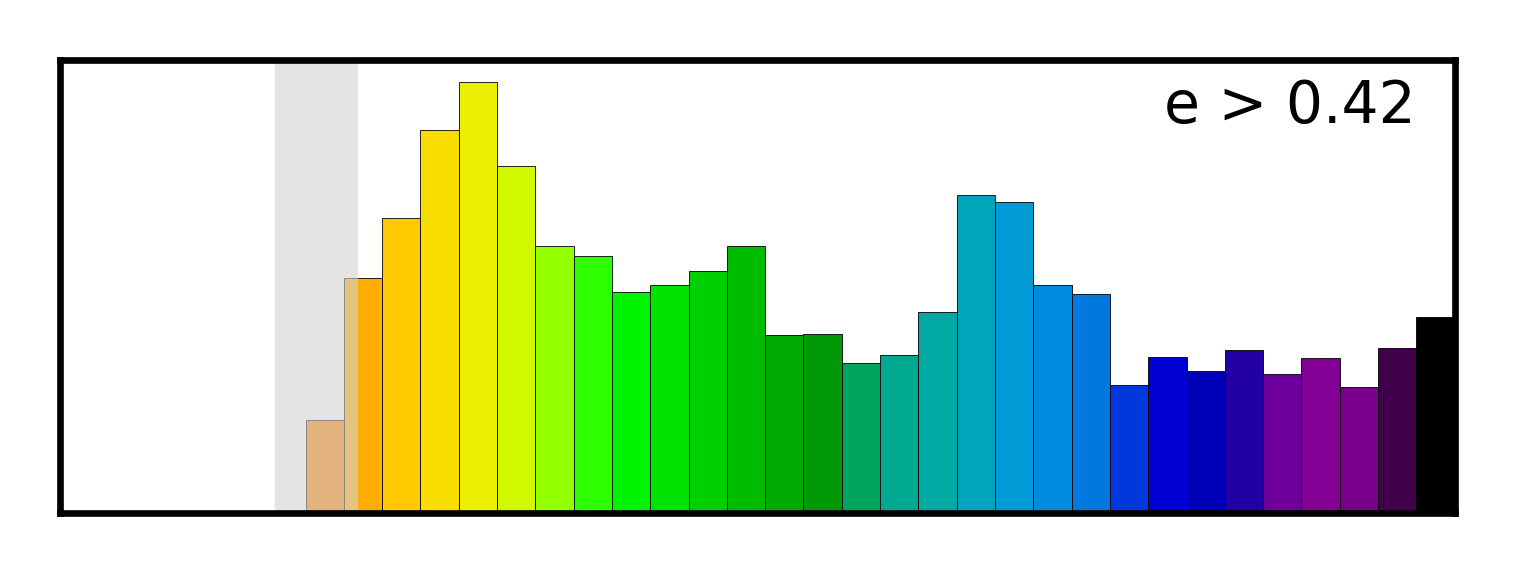}
    \includegraphics[width=1.035\textwidth]{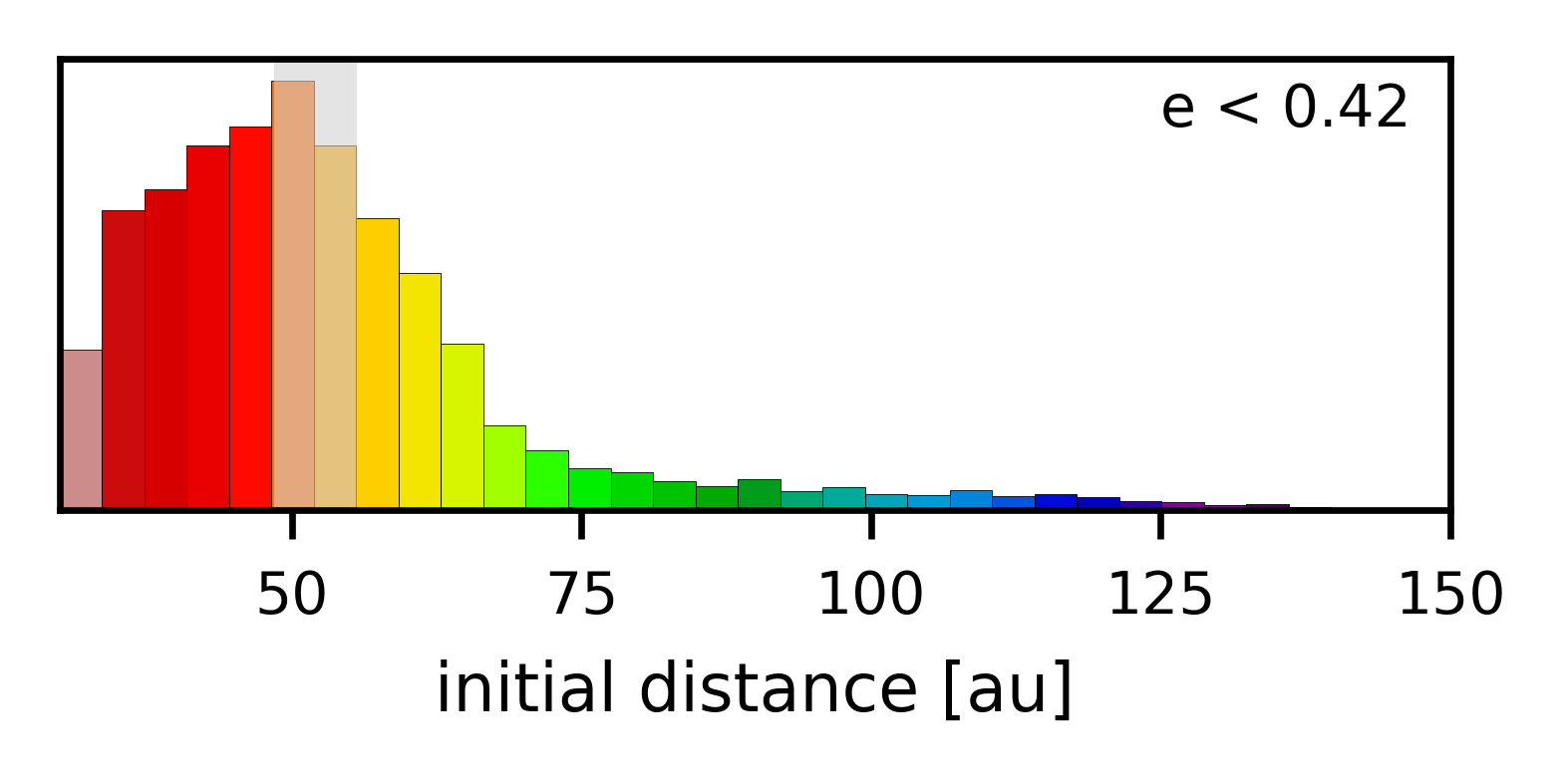}
    \textbf{(d)}
  \end{minipage}
  
\caption{{\bf Connection between TNO colour and eccentricity.} The left column illustrates the observational data of \citet{Ali:2021} and \citet{Marsset:2019}, and the right column shows the numerical results from our flyby simulation. Panels (a) and (b) display scatter plots of the TNOs' eccentricity as a function of periastron distance. Panels (c) and (d) show the corresponding colour distributions distinguishing between objects with $e >$~0.42 and $e <$~0.42. The colours in panel (b) and (d) are as defined in Fig.~\ref{fig:flyby}~(a). For the data of \citet{Marsset:2019} we excluded the centaurs.}
\label{fig:compare_ecc}
\end{figure*}

\subsection{Eccentricity correlation}
\label{sec:ecc}

We examine the correlation between the colours of TNOs and their orbital eccentricities. \citet{Ali:2021} found a lack of very red TNOs for those with $e >$~0.42 (Fig.~\ref{fig:compare_ecc}~(a)). This scarcity is clearly visible when comparing the spectral index distributions for TNOs with $e >$~0.42 (Fig.~\ref{fig:compare_ecc}~(c), top) with those for TNOs with $e <$~0.42 (Fig.~\ref{fig:compare_ecc}~(c), bottom).

Our simulations (Figs.~\ref{fig:compare_ecc}~(b) and (d)) mirror this trend, showing few red particles at high eccentricities ($e >$~0.42). Figure~\ref{fig:flyby_entire}~(d) of the Appendix illustrates that this match is also obtained for a 100~au-sized disc. High-eccentricity TNOs originate primarily from the outer disc, injected into high-inclination orbits, while low-eccentricity TNOs remain close to their origins. Thus, the flyby model simultaneously explains the scarcity of red objects at high inclination \emph{and} at high eccentricity.  We also conclude that the orbits of the few non-red TNOs among the low-eccentricity TNOs result from re-circularisation due to the long-term evolution.

In the observation and simulation scatter plots (Figs.~\ref{fig:compare_ecc}~(a) and (b)), a decline in the periastron distance of the maximum eccentricity for red objects may be apparent, but more observational data are required to confirm this additional similarity.

In summary, the suggested flyby provides a simple explanation for the connection between colour and inclination. Its simplicity is the strong point of this model. Generally, highly inclined flybys lift the population outside the red-cold Kuiper belt region to high inclinations and eccentricities. Given the complexity of the colour redistribution (Fig.~\ref{fig:flyby}~(b)), the more subtle features are sensitive to flyby parameters. Not every flyby leads to the observed colour pattern. For example, closer flybys or a higher perturber mass would lift the test particles of the cold Kuiper belt region to high inclination and eccentricities mixing of red and blue colours in the Kuiper belt region. Less inclined flybys would lead to less lifting out of the plane. The clear division would disappear in both cases and shift to other locations.

\subsection{Colours of dynamic groups}
\label{sec:groups}

The fraction of red TNOs varies among dynamic groups. Early observations indicated that the cold population is predominantly very red, whereas the hot population displays various colours \citep{Tegler:2000,Hainaut:2002,Doressoundiram:2005,Peixinho:2008}. According to \citet{Marsset:2019}, only 8~\% of classical TNOs are very red, compared to 36~\% of resonant TNOs. However, the orbits of resonant and scattered disc objects likely changed in the past by interacting with Neptune.

We concentrated on populations least influenced by the interactions with Neptune. Detached objects, defined as non-resonant TNOs with $a >$~47.4~au and $e >$~0.24 \citep{Gladman:2021}, are such a group. \citet{Marsset:2019} observed that only two of the 13 detached TNOs are very red, leading to a rate of $<$~10~\% for such objects (excluding the centaur 2007~JK43). In the DES sample, \citet{Bernardinelli:2025} found radial stratification within the NIRB population, in that hot Kuiper belt objects NIRFs are on average redder than detached NIRBs. Our flyby simulation reproduces the scarcity of red objects among detached TNOs, which is explained by detached objects primarily originating from the 55~au to 70~au region. We also replicate the trend of increasing orbital inclinations from light red to grey in the colour distribution identified by \citet{Marsset:2023}. Thus, our simulations also recover trends among subgroups of the sample, summarised in Fig.~\ref{fig:detached} in the Appendix.

\begin{figure*}
\centering

  \begin{minipage}[b]{0.40\textwidth}
    \centering
    \includegraphics[width=\textwidth]{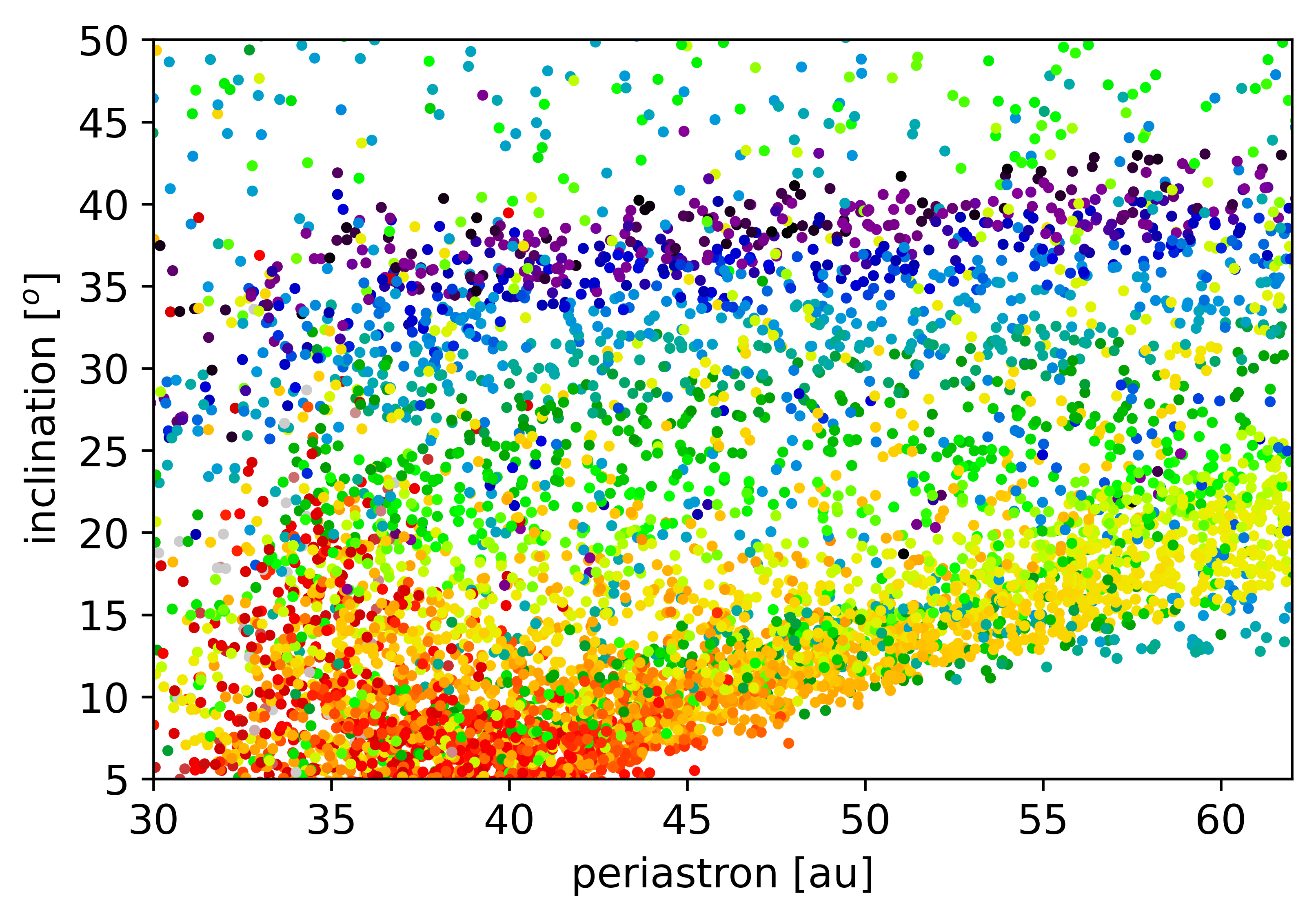}
    \textbf{(a)}
  \end{minipage}
  \begin{minipage}[b]{0.40\textwidth}
    \centering
    \includegraphics[width=\textwidth]{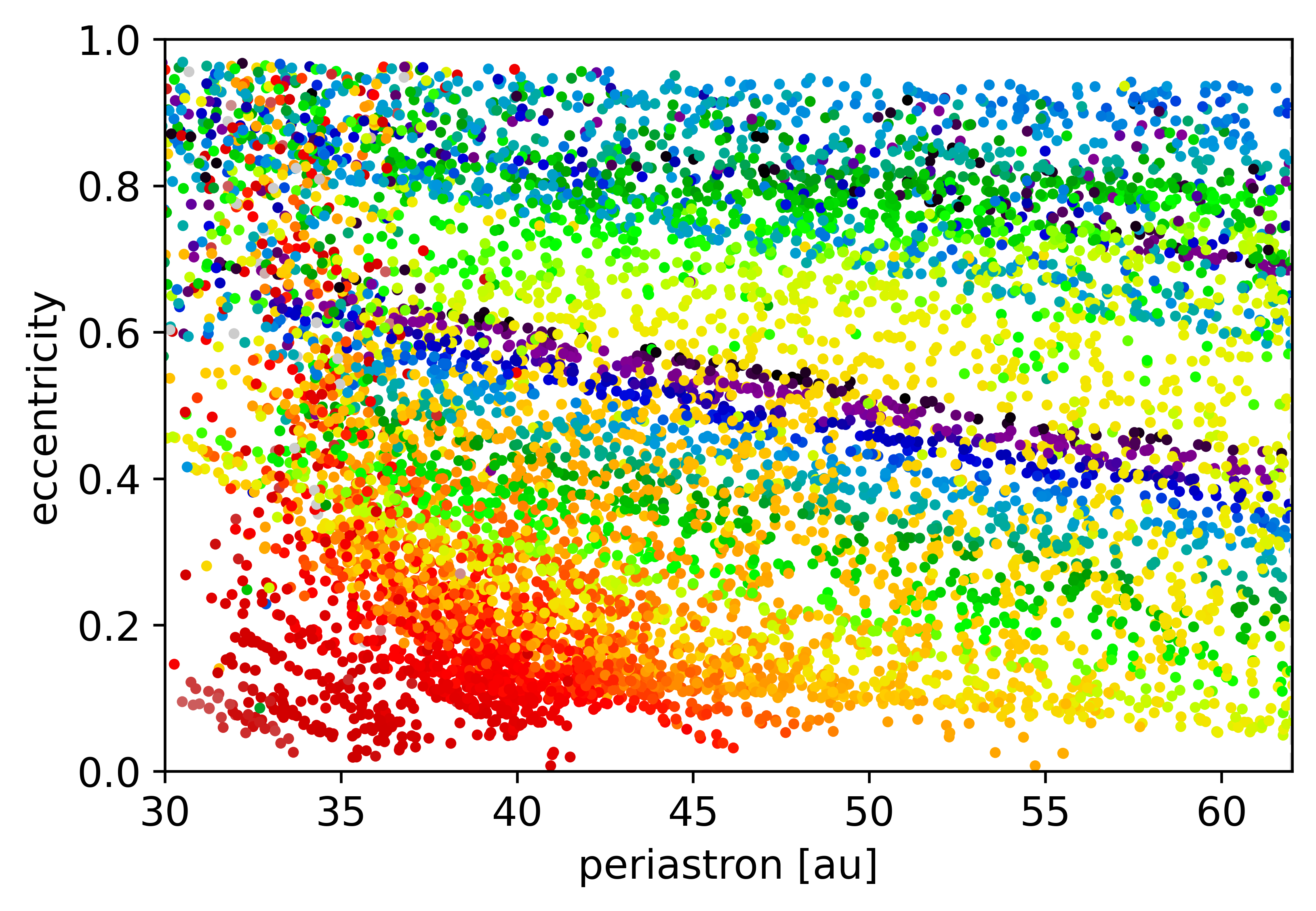}
    \textbf{(b)}
  \end{minipage}

  \begin{minipage}[b]{0.40\textwidth}
    \centering
    \includegraphics[width=\textwidth]{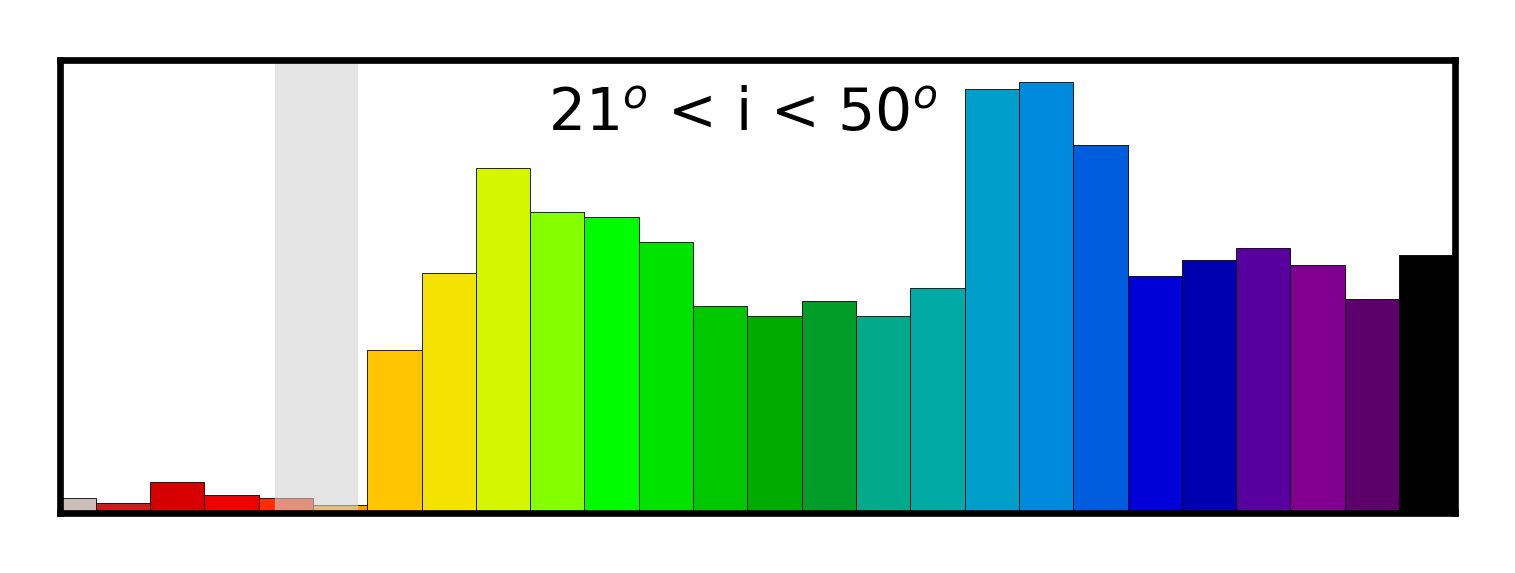}
    \includegraphics[width=1.035\textwidth]{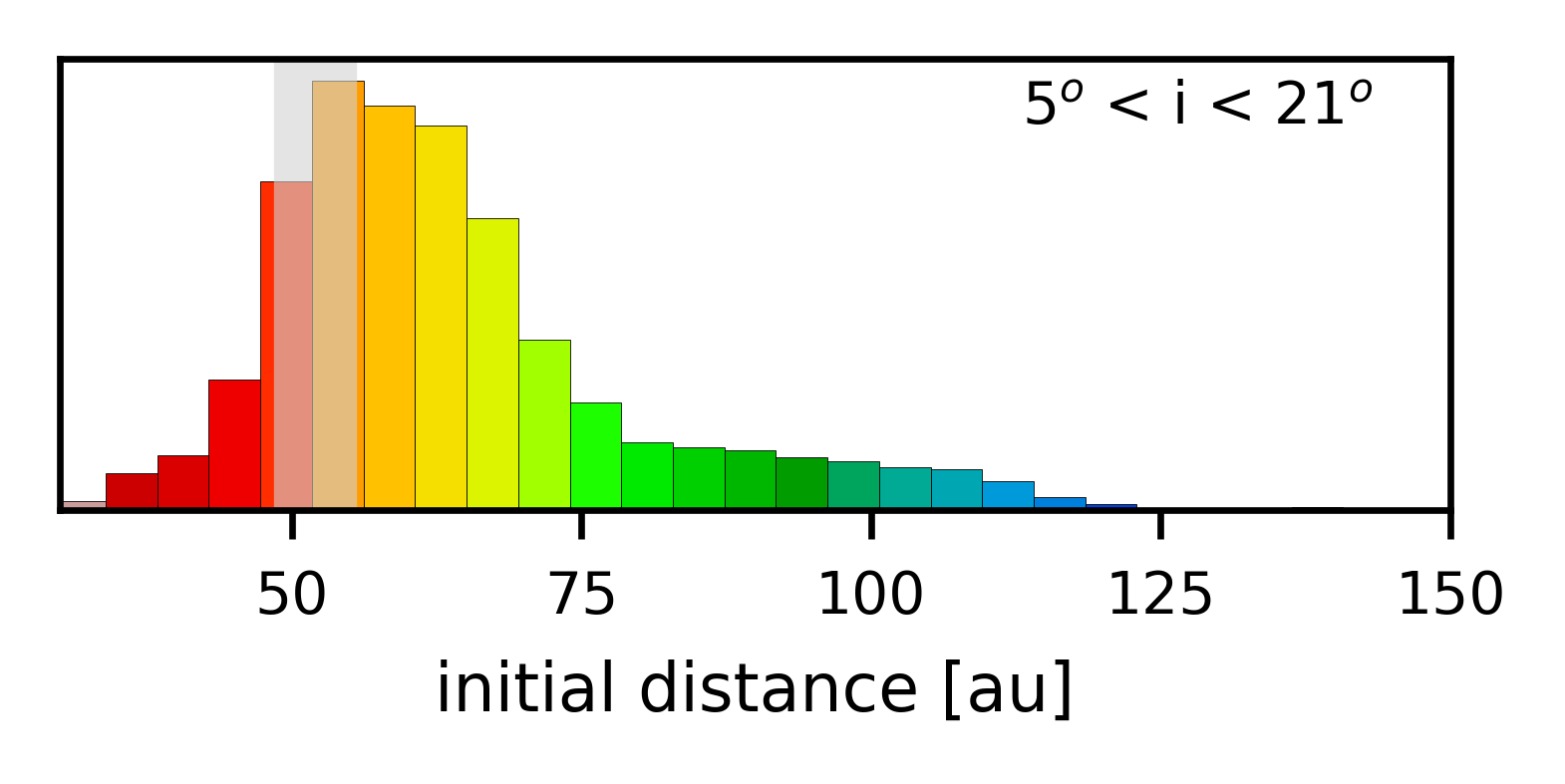}
    \textbf{(c)}
  \end{minipage}
  \begin{minipage}[b]{0.40\textwidth}
    \centering
    \includegraphics[width=\textwidth]{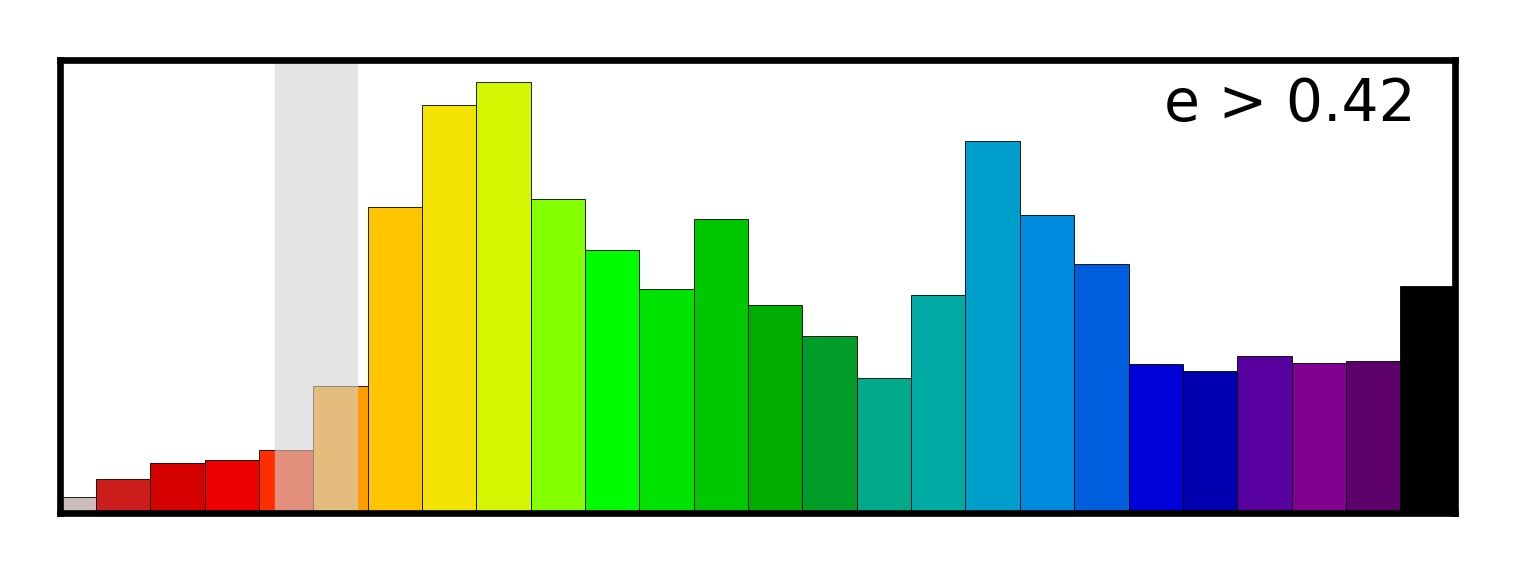}
    \includegraphics[width=1.035\textwidth]{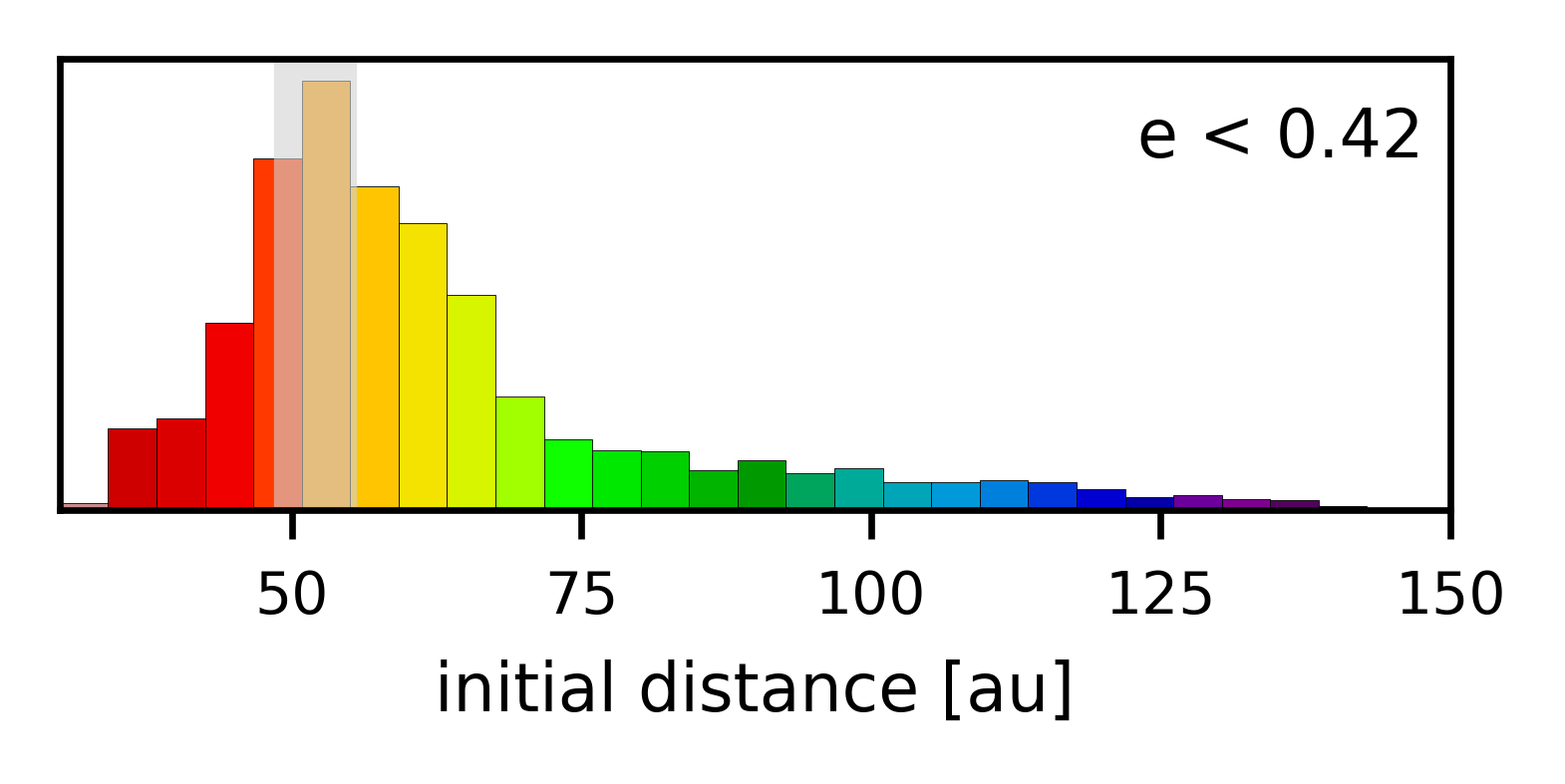}
    \textbf{(d)}
  \end{minipage}

\caption{{\bf Long-term evolution.} Connection between TNOs' colours, inclinations (a), and eccentricities (b) after 1~Gyr. Panels (c) and (d) depict the corresponding colour distributions.}
\label{fig:evo}
\end{figure*}

\subsection{Long-term evolution}
\label{sec:evo}

Interactions with the planets have affected the orbits of some TNOs, particularly those close to Neptune. We only modelled the first Gyr post-flyby due to the computational demands of tracking tens of thousands of test particles over long periods.

After 1~Gyr, the overall structure is similar, with very red objects remaining rare among high-inclination and high-eccentricity TNOs. However, the colour patterns become less distinct (Figs.~\ref{fig:evo}~(a) and (b)), as some red particles shift to high-inclination orbits or are ejected, especially those with 30~au~$< p <$~40~au, low inclinations, and low eccentricities.

The distinct differences in the colour distributions between low- and high-inclination, as well as low- and high-eccentricity, TNOs persist (Figs.~\ref{fig:evo}~(c) and (d)). The alignment between observations and simulations is maintained even after this long period. We find that interactions with Neptune are most intense in the first 100~Myr, with diminished effects afterwards, indicating that the colour patterns will persist over even longer timescales. On the dynamic side, the most significant effect of the long-term evolution is the emergence of TNOs on resonant orbits.

\begin{figure*}
\centering

  \begin{minipage}[b]{0.45\textwidth}
    \centering
    \includegraphics[width=\textwidth]{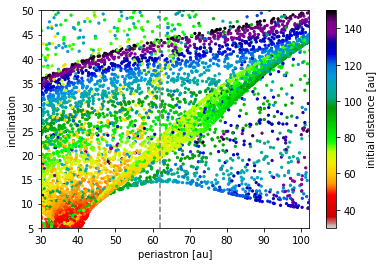}
    \textbf{(a)}
  \end{minipage}
  \begin{minipage}[b]{0.45\textwidth}
    \centering
    \includegraphics[width=\textwidth]{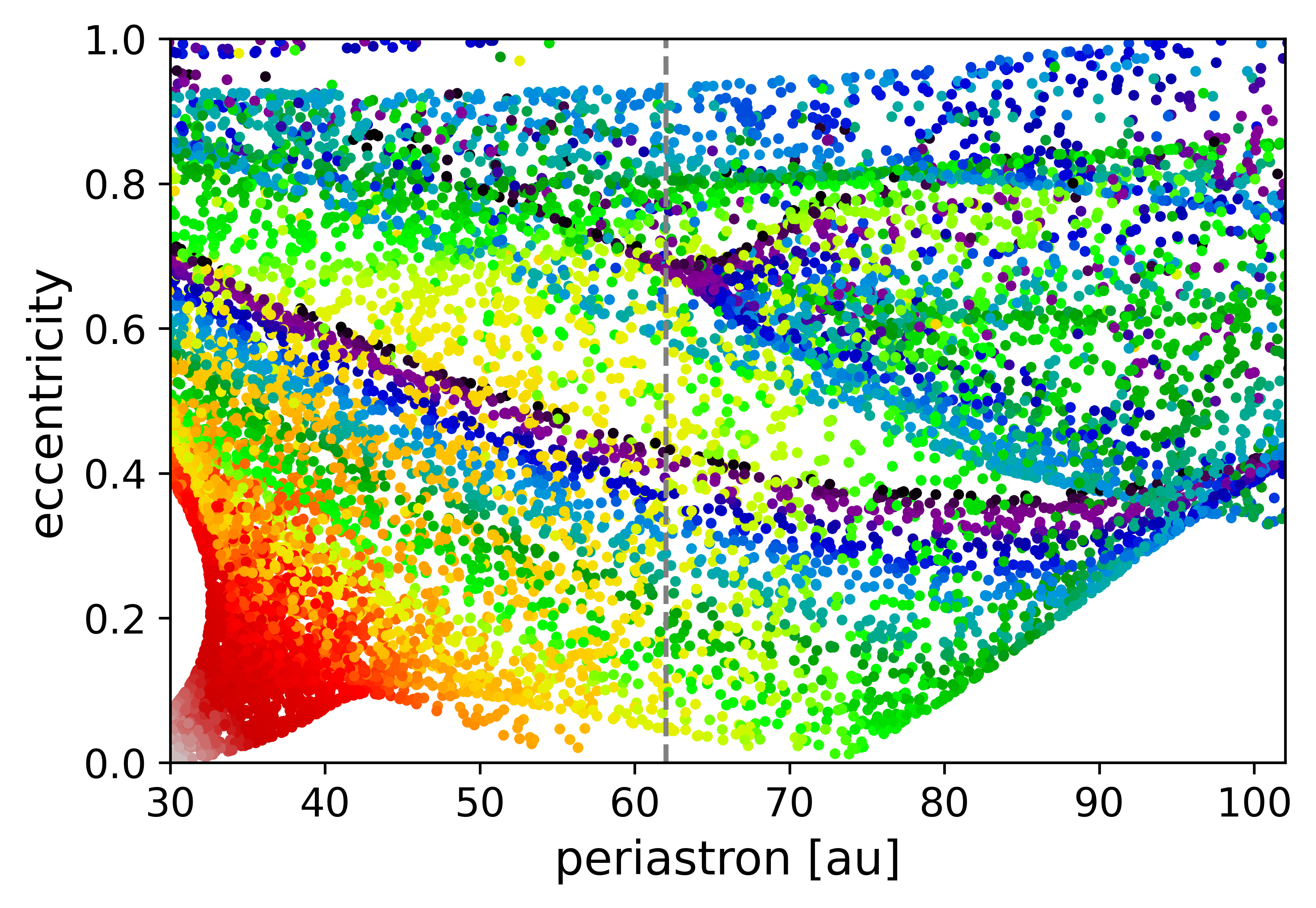}
    \textbf{(b)}
  \end{minipage}

  \begin{minipage}[b]{0.45\textwidth}
    \centering
    \includegraphics[width=\textwidth]{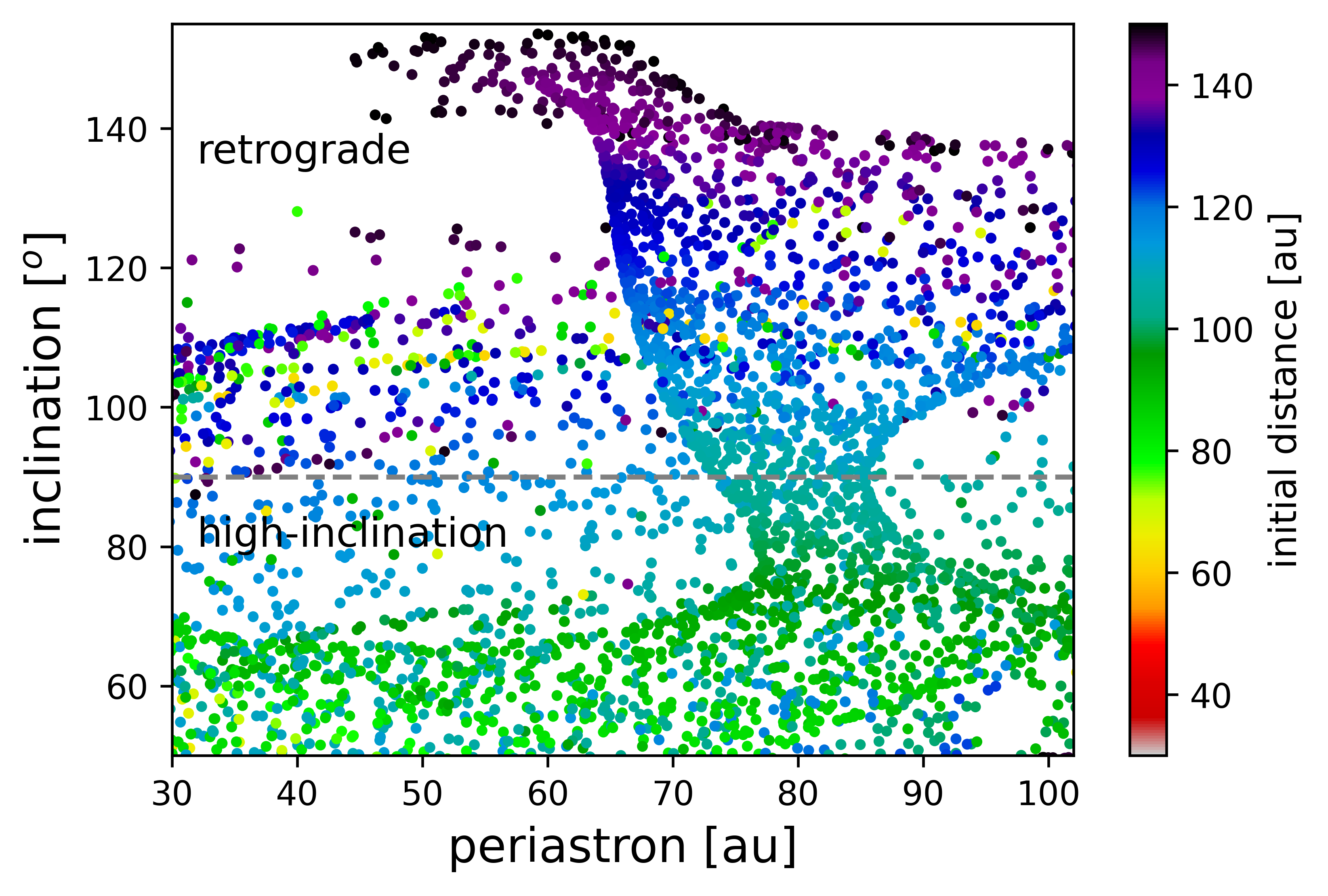}
    \textbf{(c)}
  \end{minipage}
  \begin{minipage}[b]{0.425\textwidth}
    \centering
    \includegraphics[width=\textwidth]{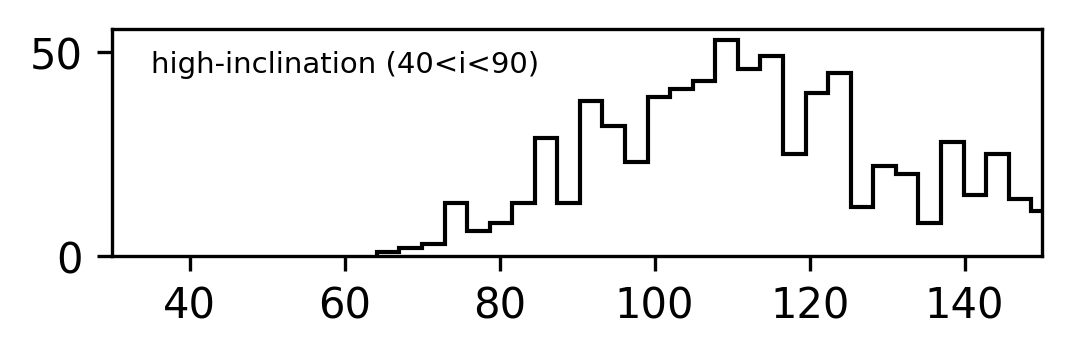}
    \includegraphics[width=\textwidth]{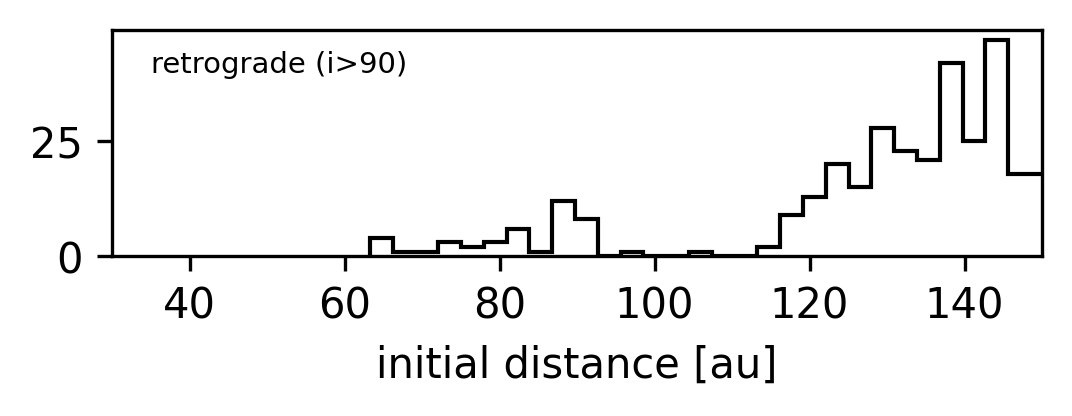}
    \textbf{(d)}
  \end{minipage}

\caption{{\bf Predictions for LSST.}  Panels (a) and (b) for TNOs with perihelion distances exceeding 60~au. Panel (c) shows the colours of highly inclined and retrograde objects with $p <$~100~au. Panel (d) illustrates the corresponding colour distributions.}
\label{fig:predict}
\end{figure*}

\subsection{Prediction for LSST discoveries}
\label{sec:predict}

LSST is projected to increase the number of detected TNOs tenfold, allowing for colour and spectral analysis. In anticipation of this, we try to predict the colours of these soon-detectable TNOs from a flyby perspective.
Focusing on distant TNOs with 60~au~$< p <$~100~au (see, Figs.~\ref{fig:predict}~(a) and (b)), we find that the test particle colours range from yellow to dark blue, with very few red objects, regardless of inclination or eccentricity. This translates into distant TNOs predominantly being light red to various shades of grey, and a notable lack of very red objects, even among low-inclination and low-eccentricity TNOs. These TNOs (60~au~$< p <$~100~au) are mainly transported inwards from regions between 65~au and 150~au.

We also predict that the colour distribution of retrograde TNOs differs significantly from prograde TNOs (Figs.~\ref{fig:predict}~(c) and (d)). They even differ from the high-inclination prograde ones (60~°~$< i <$~90~°), which are mostly medium grey, since they originate from 80~au to 110~au. In contrast, retrograde TNOs originate from \emph{two} distinct regions (70~au to 90~au and 120~au to 150~au), showing a bimodal distribution of light red and predominantly dark grey to blue-grey colours. The retrograde TNOs lack the colours most common for high-inclination TNOs. Thus, the colours of the retrograde TNOs could provide an even more rigorous test of this model and other hypotheses.

\subsection{Connection to irregular moons}
\label{sec:moons}

A further argument for a stellar flyby is based on the colours of the irregular moons. It has long been speculated that irregular satellites originate from the same planetesimal reservoir as the TNOs, possibly being captured from regions beyond Neptune shortly after the giant planets formed \citep{Jewitt:2007}. A capture would explain the distinct characteristics of irregular moons, such as orbiting their host planets in more distant, inclined, and eccentric orbits than regular moons. However, there exists one difference to the TNOs -- the irregular lack of very red objects.

Several hypotheses exist about how the TNOs moved from beyond Neptune into the region of the giant planets \citep{Nesvorny:2014}. We suggested recently that a stellar flyby could inject TNOs into the vicinity of the giants \citep{Pfalzner:2024b}. For our flyby parameters, all injected TNOs come from beyond 60~au, explaining the observed lack of very red objects (see, Fig.~\ref{fig:Moons} in the Appendix). Re-analysing the injected TNOs, we find that their colours should resemble those of high-inclination TNOs shown in Fig.~\ref{fig:compare_inc}. Thus, the stellar flyby could explain both the TNOs’ colour distribution \emph{and} the scarcity of highly red objects among the irregular moons.

\section{Discussion}
\label{sec:discuss}

\subsection{Selection effects and colour evolution}
\label{sec:biases}

Comparing simulations and observations is influenced by several biases. The faintness of TNOs limits detections to larger and brighter objects \citep{Bannister:2018}. Additionally, TNOs with large perihelion distances and those on highly eccentric or inclined orbits are often underrepresented \citep[e.g., ][]{Petit:2017}. If TNO colours are related to size or orbital parameters, these biases impact the comparison between observations and models.

We showed that for our flyby scenario a primordial disc dominated by very red bodies in the cold Kuiper belt region and objects of various shades of grey beyond 50~au can reproduce the observed colour. However, it is unclear how this is related to the chemical composition of the primordial disc. There are two reasons for this uncertainty: First, the link between the chemical compositions of TNOs and their colours is not fully understood. Second, it is unclear to what extent the surface differs from the bulk compositions of TNOs. Factors such as UV irradiation, cosmic rays, and particle impacts may have altered the surfaces of TNOs \citep{Brown:2011}. Unfortunately, the chemical distribution in observed discs cannot solve this problem. Theoretically, disc stratification is believed to be the result of the position of snowlines. However, actual disc observations reveal considerable variation in material distribution within discs \citep{Oeberg:2023}, which challenges the idea that snowlines are the only factor influencing a disc's chemical structure \citep{Perotti:2023}.
The complexity of TNO and disc chemistry makes it difficult to identify the dominant processes that could have led to a red-to-grey stratification in the Solar System's disc.

For irregular moons, \added{the situation is even more complex. Outgassing of} volatile materials \citep{Jewitt:2007,Jewitt:2018} or collisional break-up \citep{Ashton:2021} may have altered their appearance in addition.

\subsection{Comparison to planet instability model}
\label{sec:nice}

Several other models attempt to explain the dynamics of TNOs \citep[e.g., ][]{Jilkova:2015,Vokrouhlinky:2024,Brown:2024}, but only the planet instability model has explored the connection between TNO dynamics and their colours to date. This model posits that TNOs originate from the region between the giant planets, were ejected due to planet migration, and some were recaptured in the outer Solar System \citep[e.g., ][]{Fernandez:1984, Hahn:1999, Morbi:2003, Levison:2008, Raymond:2018}. While this effectively accounts for classical Kuiper belt objects, it requires additional mechanisms to account for TNOs with retrograde or Sedna-like orbits.

Studies on TNO colours \citep{Nesvorny:2020, Ali:2021} suggest a grey population close to the Sun and a red population farther out. The model indicates that the planet instability alters the TNO orbits, mixing these populations, with the resulting colour distribution depending on the initial colour divide and the type of Neptune's migration \citep[e.g., ][]{Brown:2011,Nesvorny:2020}. \citet{Ali:2021} found colour transitions at 38~au or 42~au could explain the scarcity of very red TNOs at high inclinations. However, a 38~au transition overrepresents very red scattered disc objects, while a 42~au transition underrepresents very red plutinos, indicating that the link between TNO dynamics and colours remains unresolved.

\section{Conclusion}
\label{sec:conclude}

We investigated whether the close flyby of a perturbing star (\mbox{$M_P =$~0.8~\MSun}, \mbox{$r_P =$~110~au}, \mbox{$i_P =$~70~°}) could explain the complex colour distribution of the TNOs. Assuming an initial colour gradient in the Sun's debris disc, we found that the flyby accounts for the observed colour correlations from the OSSOS and DES surveys. Specifically, those very red TNOs:
\begin{itemize}
    \item become increasingly scarce for higher inclinations ($i >$~21~°),
    \item are rare among very eccentric ($e >$~0.42) TNOs,
    \item and within the detached population.
\end{itemize}
We also reproduce the trend of increasing inclinations towards the bluer end of the colour distribution of bright infrared object. We connect it to the transport in spiral arms created during the flyby. The long-term evolution over 1~Gyr changes the colour structure only slightly.

This simultaneous explanation of the TNO dynamics and colours significantly strengthens the argument for a stellar flyby largely determining the structure of the Solar System beyond Neptune. Moreover, the dynamics and colours of the irregular moons can also be accounted for \citep{Pfalzner:2024b}.

Upcoming instruments like LSST are expected to detect tens of thousands of new TNOs. We predict that (1) distant TNOs will mainly be grey to blue-grey, unless they are dwarf planets, (2) retrograde TNOs will show a bimodal colour distribution, with some light red objects and a distinct blue-grey group, but lacking the colour most common among high-inclination TNOs, and (3) the inclination vs. periastron correlation will reveal stripes of blue-grey TNOs at high inclinations. These predictions can be tested within three years, potentially confirming the flyby hypothesis and clarifying its influence on the outer Solar System.

\begin{acknowledgments}
The authors thank Pedro Bernardinelli for the valuable discussions on the DES results. We would also like to thank the referee for the valuable comments and suggestions, which have helped to improve the quality of the manuscript. Our appreciation further extends to the editor for the efficient handling of our submission. The simulations were performed on the JUWELS and JURECA supercomputers at the Jülich Supercomputing Center.
\end{acknowledgments}

\begin{contribution}
Conceptualisation, S.P.; Simulation of the flyby and long-term evolution, F.W.; Diagnostic: S.P., P.G.; Comparison to observational data, S.P.; Writing, S.P., F.W., P.G.; Data curation, F.W.


\end{contribution}



\appendix

\section{Entire Parameter space and dependence on disc size}

In the main part of the paper, we concentrate on the parameter space accessible by observations. The complete parameter space showing the resulting correlation between TNOs' colours and their dynamical parameters is illustrated in Figs.~\ref{fig:flyby_entire}~(a) and (b). Panels (a) and (b) show the inclinations and eccentricities of the TNOs as functions of periastron distance, respectively. The symbol colours indicate the TNOs' origin in the pre-flyby disc. The red colour corresponds to the red TNOs, whereas the other colours represent the shades of grey observed for the TNOs. The red TNOs concentrate at low inclinations, eccentricities, and periastron distances. This reflects the fact that they retain much more of their original dynamics than particles at larger distances from the Sun.

The emerging pattern reflects the two primary spiral arms that form during the flyby, along with some less pronounced higher-order arms. In the spiral arms, two adjacent streams of planetesimals with different velocities form -- one with sub-, the other with super-Keplerian velocities.
The sub-Keplerian stream moves particles inwards, while the super-Keplerian stream pushes planetesimals outwards \citep{Pfalzner:2003}. 
An example of this effect can be seen in the dark blue stripes of the inclination-periastron plot (see, Fig.~\ref{fig:flyby_entire}~(a)). These particles were originally located beyond 100~au and have moved inwards along the edges of the spiral arms.

A large unknown is the disc size before the flyby. Here we show the result for up to 150~au-sized discs. The rainbow colour plate can be used to illustrate the result for smaller discs. Figures~\ref{fig:flyby_entire}~(c) and (d) show the results for the example of a 100~au pre-flyby disc. The parameter space would lack the blue particles in the scatter plots of Fig.~\ref{fig:compare_inc}~(b) and Fig.~\ref{fig:compare_ecc}~(b). Equivalently, the plots of the distributions would simply stop at 100~au. The general result of the lack of very red objects among high-inclination and high-eccentricity TNOs is largely insensitive to disc size as long as it exceeds 70~au.

\renewcommand\thefigure{A\arabic{figure}}
\setcounter{figure}{0}

\begin{figure*}[ht]
  \centering
  
  \begin{minipage}[b]{0.45\textwidth}
    \centering
    \includegraphics[width=\textwidth]{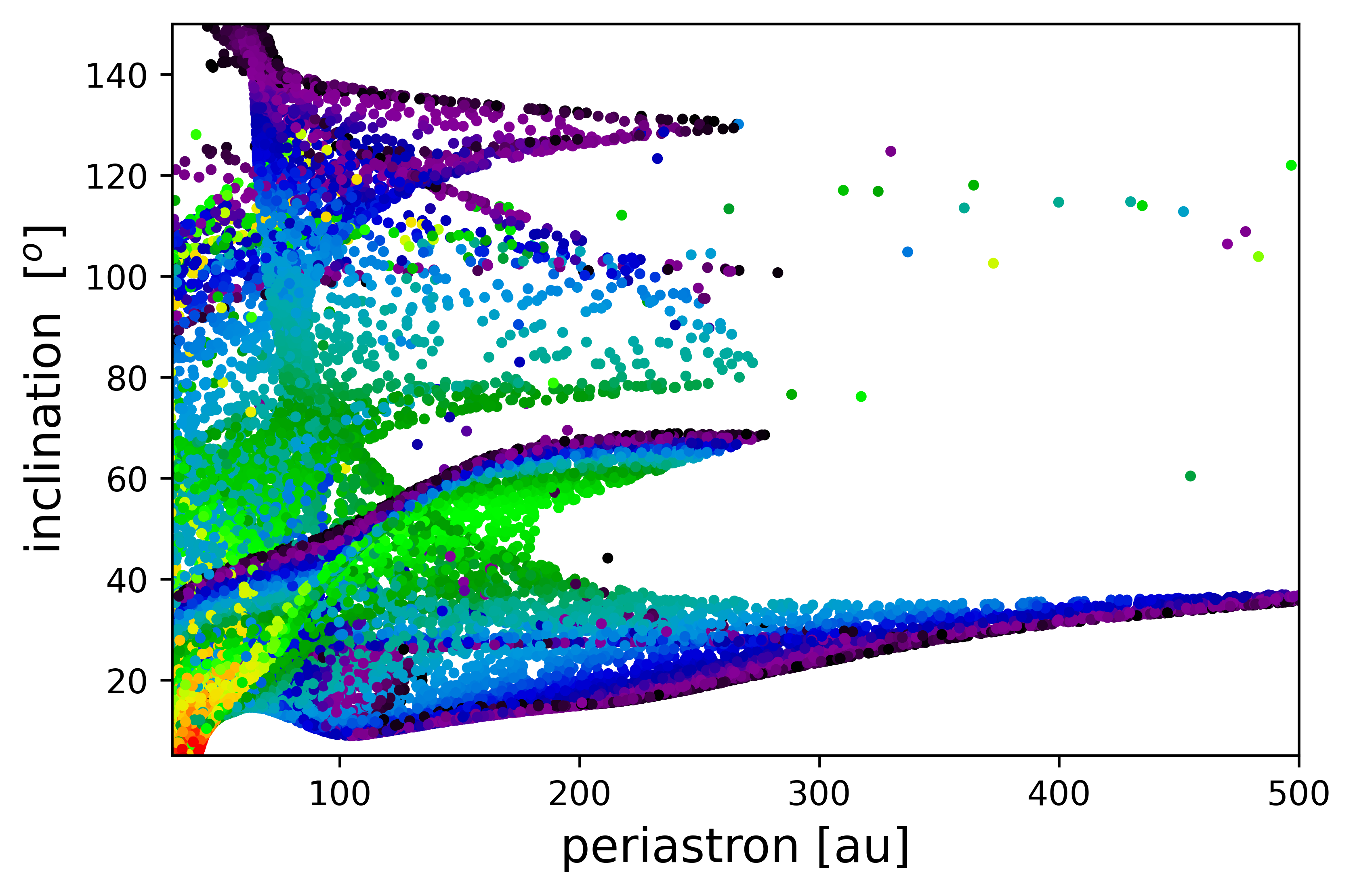}
    \textbf{(a)}
  \end{minipage}
  \begin{minipage}[b]{0.45\textwidth}
    \centering
    \includegraphics[width=\textwidth]{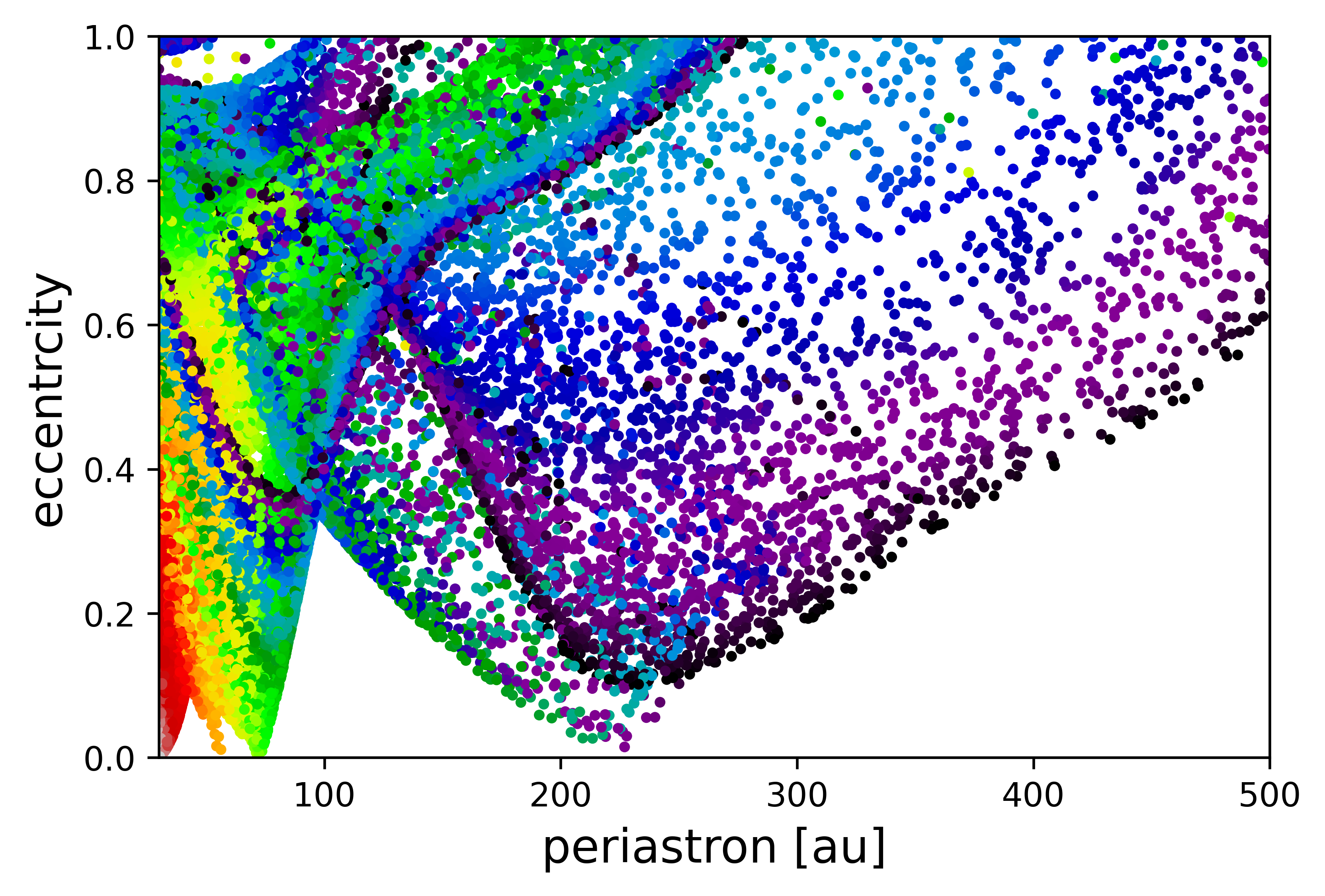}
    \textbf{(b)}
  \end{minipage}

  \begin{minipage}[b]{0.45\textwidth}
    \centering
    \includegraphics[width=\textwidth]{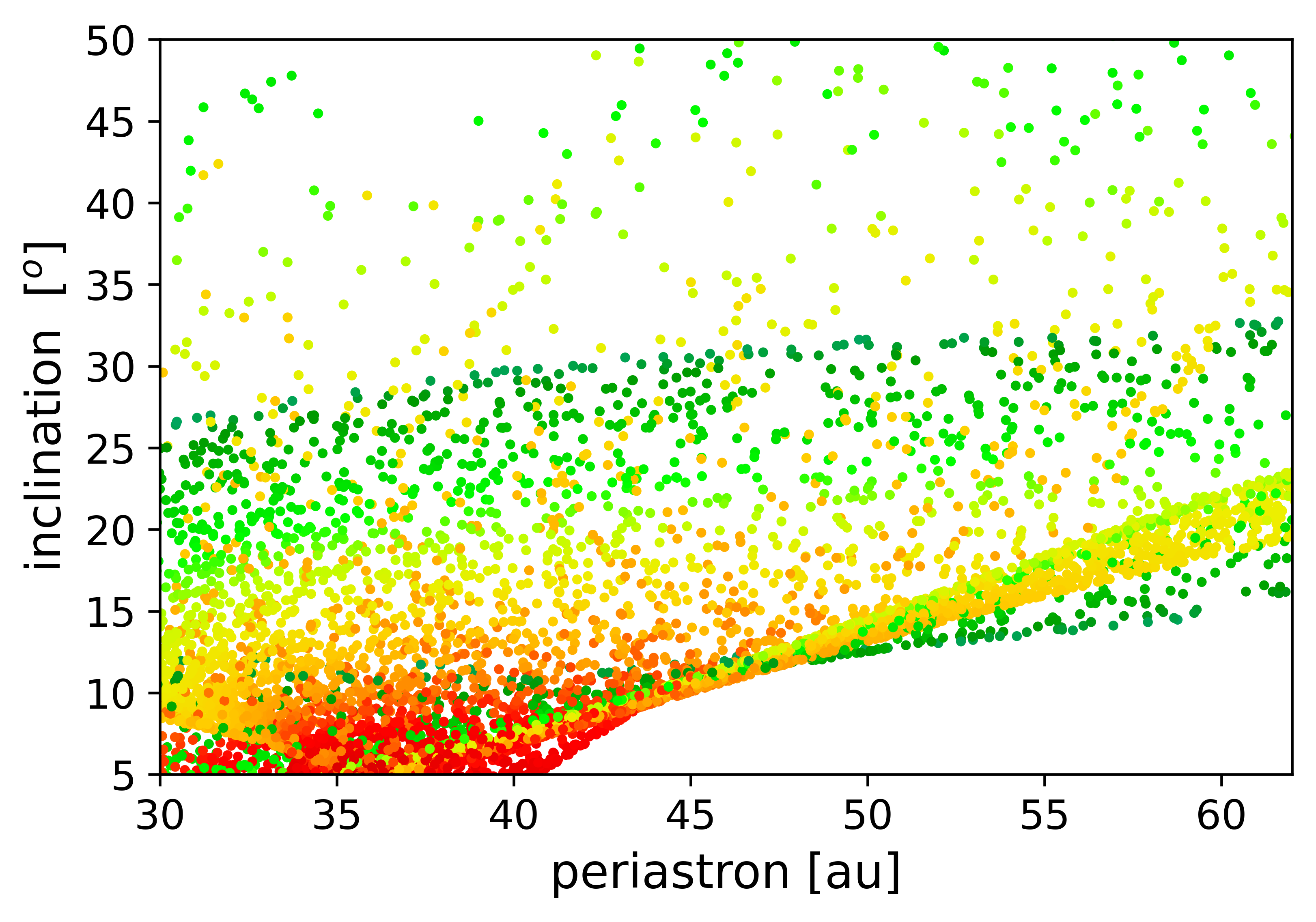}
    \textbf{(c)}
  \end{minipage}
  \begin{minipage}[b]{0.45\textwidth}
    \centering
    \includegraphics[width=\textwidth]{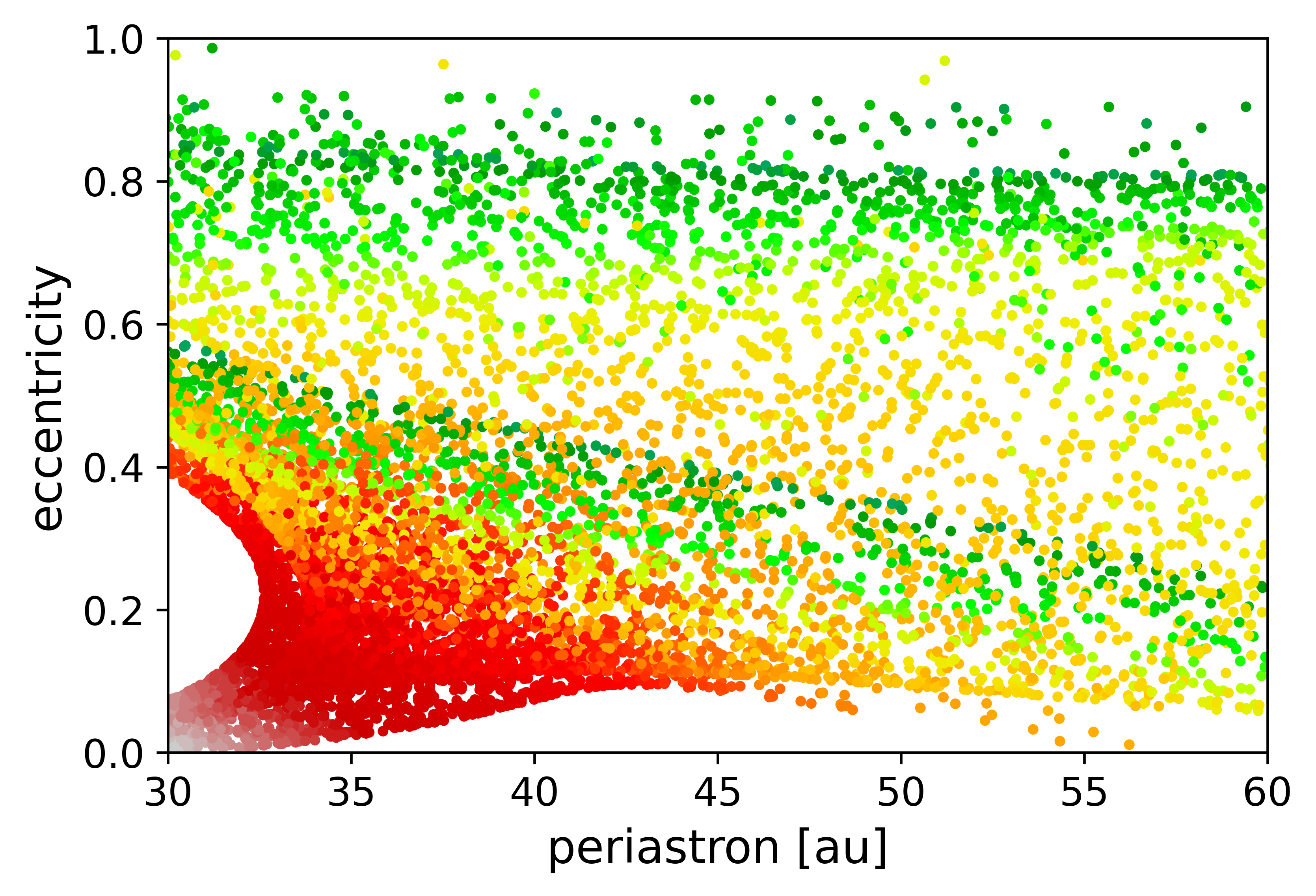}
    \textbf{(d)}
  \end{minipage}
  
  \caption{{\bf Entire parameter space and 100~au-sized disc.} Scatter plot of the TNOs' inclinations as a function of perihelion distance for (a) the entire parameter space and (c) a 100~au-sized disc. Scatter plot of the TNOs' eccentricities as a function of perihelion distance for (b) the entire parameter space and (d) a 100~au-sized disc.}
  \label{fig:flyby_entire}
\end{figure*}

\section{Dependence on surface density in pre-flyby disc}

The surface density distribution in the pre-flyby disc is unknown. In the main part of the paper, we show the test particle density. The result for a given surface density is obtained by assigning different masses to the test particles \citep{Hall:1997,Steinhausen:2012, Pfalzner:2024a}. Figure~\ref{fig:surface} compares the inclination distributions similar to Fig.~\ref{fig:compare_inc}~(d) for constant, $1/r$-dependent, and $1/r^{3/2}$-dependent initial surface densities. It shows that the general result of a relative lack of very red TNOs at high-inclination TNOs remains valid. However, the density of distant TNOs is lower for steeper surface densities.

\renewcommand\thefigure{A\arabic{figure}}
\setcounter{figure}{1}

\begin{figure*}[ht]
\centering

  \begin{minipage}[b]{0.45\textwidth}
    \centering
    \includegraphics[width=\textwidth]{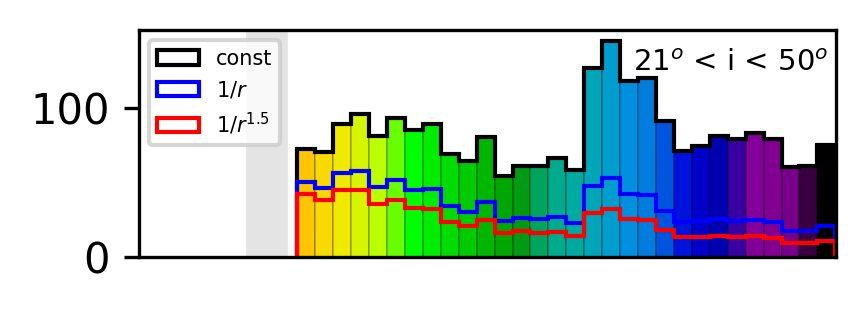}
  \end{minipage}
  
  \begin{minipage}[b]{0.475\textwidth}
    \centering
    \includegraphics[width=\textwidth]{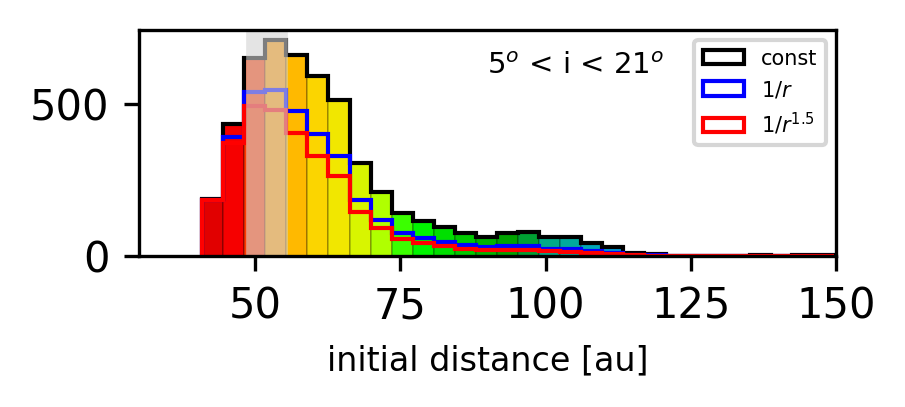}
  \end{minipage}

\caption{{\bf Surface density dependence.} Origin of the high- and low-inclination TNOs as in Fig.~\ref{fig:compare_inc}~(d). However, here we show a comparison of the test particle distribution (black), with those assuming a $1/r$ (blue) and a $1/r^{3/2}$ (red) surface distribution in the pre-flyby disc.}
\label{fig:surface}
\end{figure*}

\section{Detached TNOs}

\citet{Gladman:2021} define detached objects as non-resonant TNO with $a >$~47.4~au and $e >$~0.24. The data presented by \citet{Marsset:2019} confirmed earlier results, which hinted at a scarcity of very red objects among detached TNOs. They find that only two out of their 13 detached objects exhibit red colours (see, Fig.~\ref{fig:detached}~(a), top). One of the red objects (2007~JK43) has a periastron distance of less than 30~au, which would not be included in the diagnostics in our simulations. Thus, the rate of very red detached objects is $<$~10~\%. In the DES sample, detached TNOs are consistent with about 70~\% being grey. \citet{Bernardinelli:2025} also find evidence for radial stratification within the primordial NIRB population, with hot Kuiper belt object NIRBs being, on average, redder than detached NIRBs.

Our flyby simulation reproduces the trend of a scarcity of red objects among detached TNOs. The simulation results indicate that the detached objects originate primarily from the region 50~au to 55~au (see, Fig.~\ref{fig:detached}~(a), bottom). Consequently, grey objects dominate over red TNOs significantly in the detached population, and red objects are scarce. \citet{Marsset:2023} identified a trend of increasing orbital inclinations from the lightly red to the grey end of the colour distribution, both in the optical and near-infrared. Our simulation also recovers this trend (see, Fig.~\ref{fig:detached}~(b)).

\renewcommand\thefigure{A\arabic{figure}}
\setcounter{figure}{2}

\begin{figure*}[h]
  \centering
  
  \begin{minipage}[b]{0.40\textwidth}
    \centering
    \includegraphics[width=\textwidth]{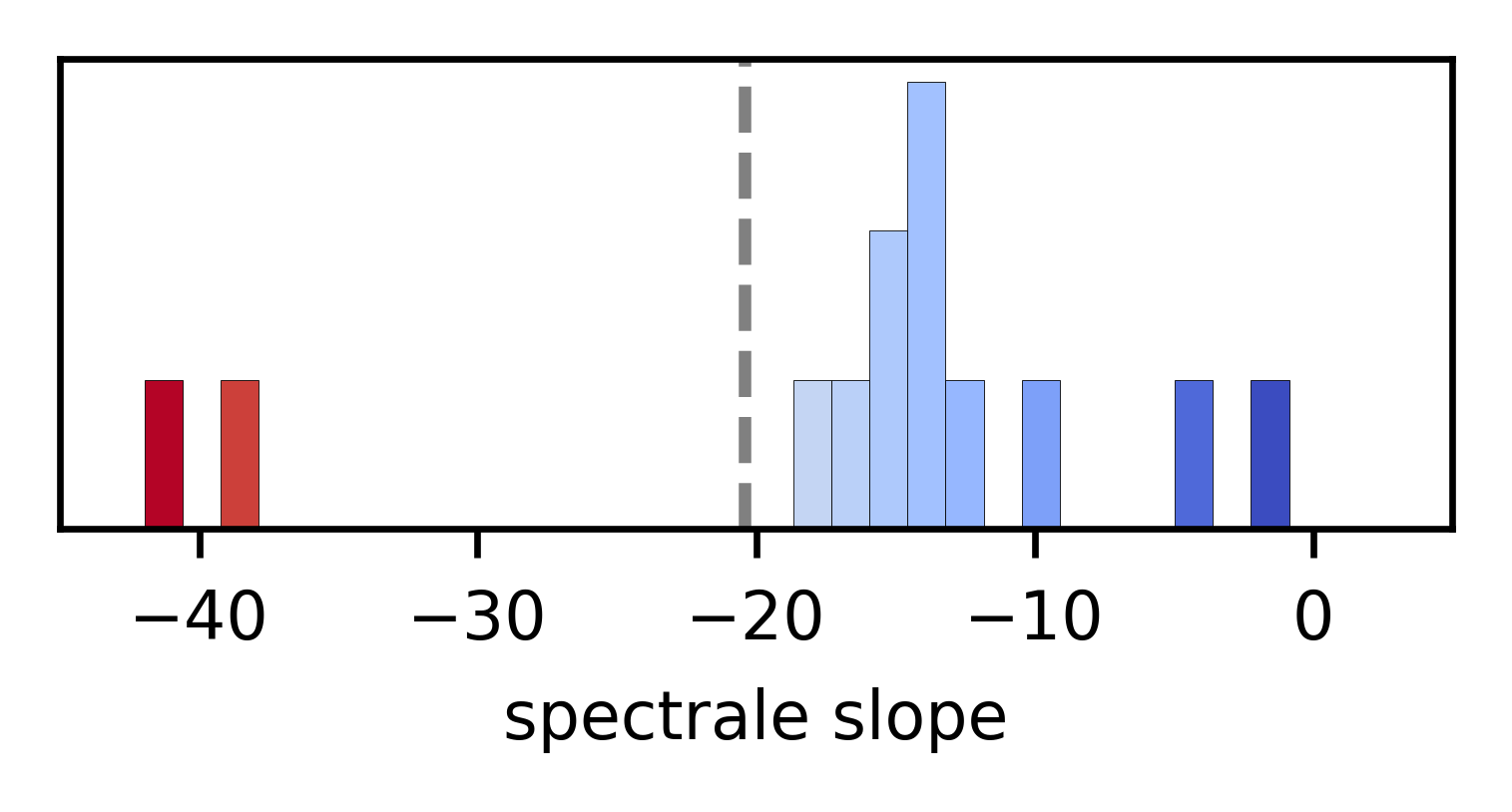}
    \includegraphics[width=1.035\textwidth]{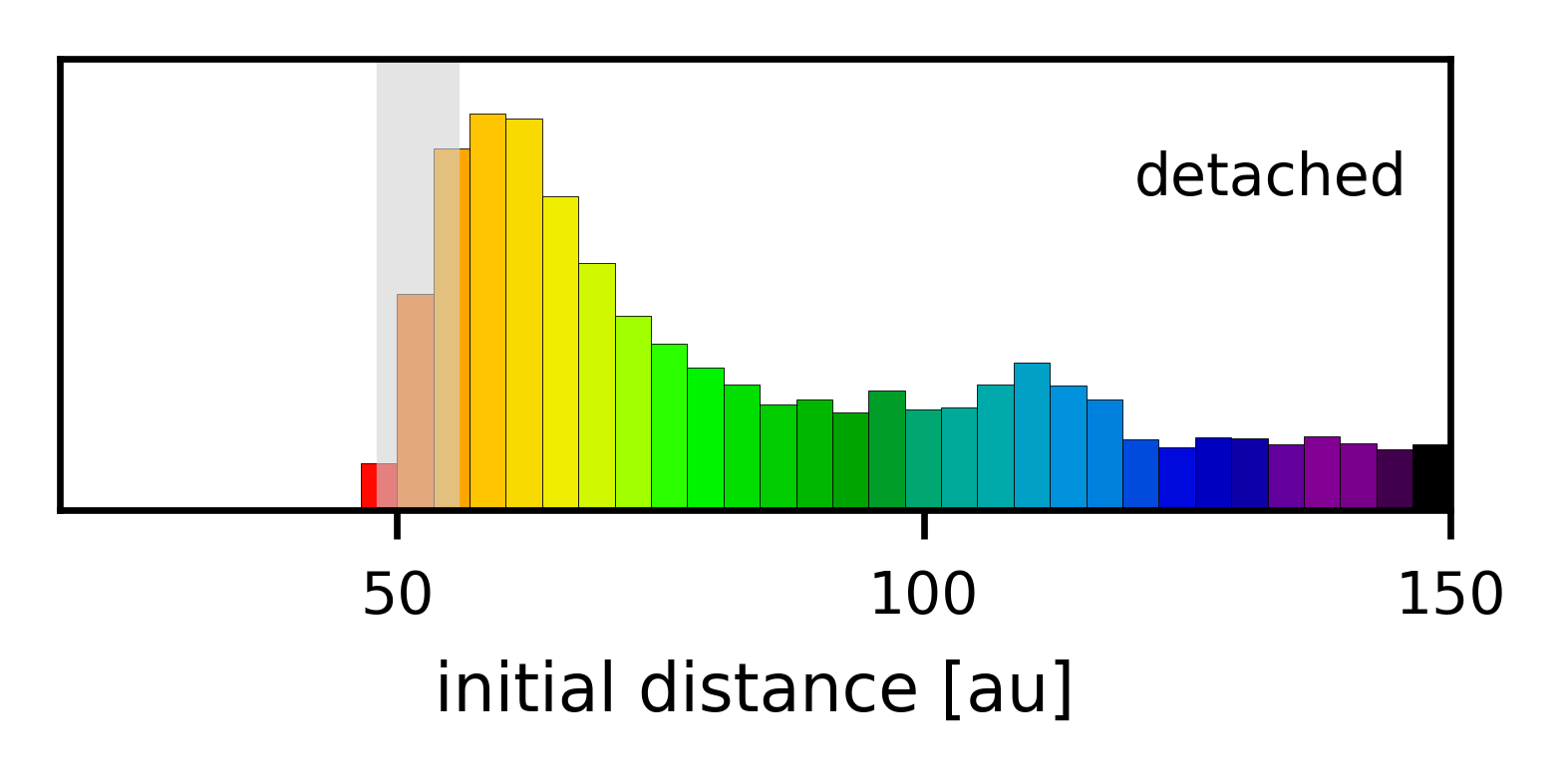}
    \textbf{(a)}
  \end{minipage}
  \begin{minipage}[b]{0.45\textwidth}
    \centering
    \includegraphics[width=\textwidth]{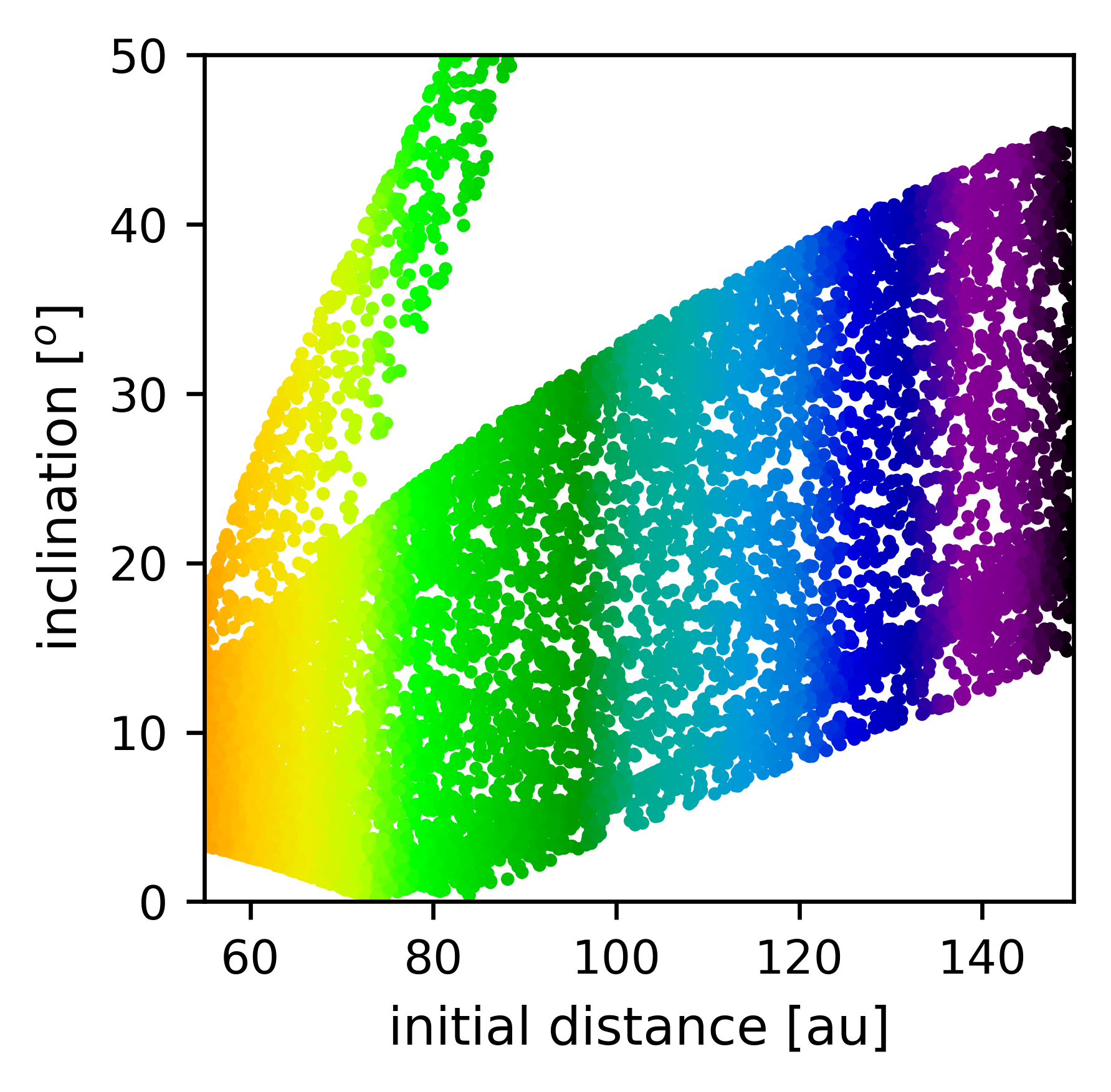}
    \textbf{(b)}
  \end{minipage}
  
  \caption{{\bf Other correlations between colours and dynamical parameters.} Panel (a) illustrates the scarcity of very red objects among the detached TNOs in the observations (top) and simulations (bottom). Panel (b) depicts the inclination as a function of the initial distance of the TNOs.}
  \label{fig:detached}
\end{figure*}

\section{Injected TNOs}

In the stellar flyby modelled here, all the injected TNOs originate from beyond 60~au. Figure~\ref{fig:Moons} shows the origin distribution of the injected TNOs. It shows the expected scarcity of very red objects. Furthermore, it resembles to a high degree the distribution we found for the high-inclination ($i >$~21~°) TNOs shown in Fig.~\ref{fig:compare_inc}. Therefore, we also expect their colour distribution to match. Consequently, the stellar flyby would not only account for the TNOs' colour distribution but also explain the observed absence of highly red objects among the irregular moons. This scarcity of very red objects would directly result from the origin of the injected TNOs in the outer areas of the disc.

\renewcommand\thefigure{A\arabic{figure}}
\setcounter{figure}{3}

\begin{figure*}[ht]
\centering
\includegraphics[width=0.40\textwidth]{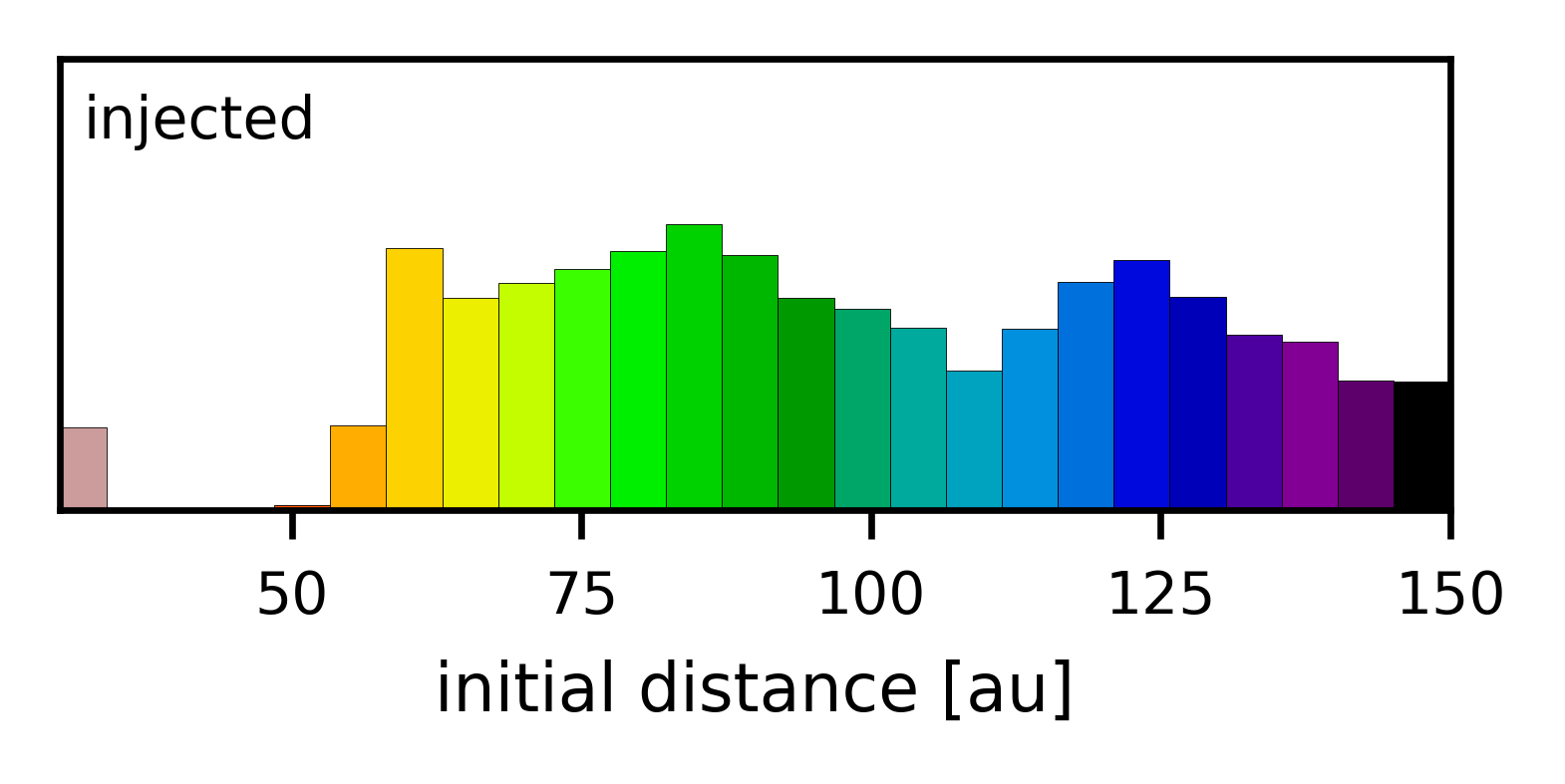}
\caption{{\bf Injected TNOs.} Origin distribution of TNOs injected within 29~au, which may have been partly captured as irregular moons.}
\label{fig:Moons}
\end{figure*}

\newpage

\bibliography{reference}{}
\bibliographystyle{aasjournalv7}



\end{document}